\documentclass[Times,twocolumn,tighten,deluxetable]{aastex702}
\usepackage{graphicx}
\usepackage{rotating}
\usepackage{lineno}
\usepackage{multirow}
\usepackage{enumerate}

\usepackage[T1]{fontenc}
\usepackage{ae,aecompl}

\accepted{November 17, 2026}
\shorttitle{STP view of GRS 1739-278}
\shortauthors{S. Srimani et al.}

\begin{document}

\title[STP View of GRS 1739-278]{\textnormal{Spectro-Temporal-Polarimetric View of GRS 1739-278 during its 2025--2026 outburst: Insights from Multi-mission Results}}

\correspondingauthor{Dipak Debnath}
\email{dipakcsp@gmail.com}

\author[0009-0004-5915-3789]{Subham Srimani}
\affiliation{Institute of Astronomy, National Tsing Hua University, Hsinchu 300044, Taiwan}
\email{subhamsrimani2023@gmail.com}

\author[0000-0003-1856-5504]{Dipak Debnath}
\affiliation{Institute of Astronomy, National Tsing Hua University, Hsinchu 300044, Taiwan}
\affiliation{Indian Center for Space Physics, 466 Barakhola, Netai Nagar, Kolkata 700099, India}
\affiliation{Center for Theory-Computation-Data Science Research (CTCD), National Tsing Hua University, Hsinchu 300044, Taiwan}
\email{dipakcsp@gmail.com}

\author[0000-0002-5617-3117]{Hsiang-Kuang Chang}
\affiliation{Institute of Astronomy, National Tsing Hua University, Hsinchu 300044, Taiwan}
\affiliation{Department of Physics, National Tsing Hua University, Hsinchu 300044, Taiwan}
\affiliation{Center for Theory-Computation-Data Science Research (CTCD), National Tsing Hua University, Hsinchu 300044, Taiwan}
\email{hkchang@mx.nthu.edu.tw}

\author[]{Anuj Nandi}
\affiliation{Space Astronomy Group, ISITE Campus, U. R. Rao Satellite Centre, Outer Ring Road, Marathahalli, Bangalore 560037, India}
\email{}

\begin{abstract}
The Galactic transient X-ray binary GRS~1739--278 was monitored with \textit{Swift}/XRT, \textit{NuSTAR}, \textit{IXPE}, and \textit{Insight-HXMT} during its recent $2025$--$2026$ outburst. We present a comprehensive spectro-temporal-polarimetric study using observations from all missions, complemented by daily monitoring with \textit{MAXI}/GSC. The outburst exhibits a fast-rise exponential-decay profile of duration $\sim97$~days without the canonical `q'-track in the Hardness--Intensity Diagram unlike previous outbursts. Timing analysis reveals transient type-B low-frequency QPOs with centroid frequencies decreasing from 0.32 to 0.23 Hz. Broadband ($1$--$40$~keV) spectral modeling around peak of the outburst shows that the emission is well described by a multicolor disk (\texttt{diskbb}), Comptonization (\texttt{thcomp}), and relativistic reflection. The spectra are dominated by thermal disk emission with disk flux exceeding the Comptonized component by nearly an order of magnitude. Reflection is better reproduced by the thermal reflection model (\texttt{relxillNS}) than the standard non-thermal model (\texttt{relxill}) indicating a thermally dominated reprocessing environment. The inferred parameters ($\Gamma\sim1.9$--$2.0$, $T_{\rm in}\sim0.9$~keV) confirm softer spectral states with thermal disk ($F_{\rm dbb}\sim49$--$76\%$) and reflection ($F_{\rm ref}\sim24$--$48\%$) dominating over non-thermal emission ($F_{\rm nth}\sim1$--$3\%$).
Model independent analyzes yield a polarization degree of $\sim2.28$–$2.43\%$ at a polarization angle of $62.0^\circ$–$66.1^\circ$ in the $2$–$8$ keV band, rising to $10.2$–$13.3\%$ in the $6$–$8$ keV band despite minimal non-thermal emission. Relativistic reflection modeling constrains the black hole spin to 
$a_k\sim0.95$--$0.98$ and disk inclination to $58^\circ$–$64^\circ$, while spectro-polarimetry reveals contributions from both disk and returning reflected photons, probing the inner accretion flow geometry.

\end{abstract}


\section{Introduction}

Black hole X-ray binaries (BH-XRBs) are among the most important astrophysical systems for studying accretion physics under the influence of strong gravitational fields. In these systems, matter from a companion star is transferred to the black hole via Roche lobe overflow through the inner Lagrange point (L1), forming an accretion disk around the compact object \citep{SS73, NT73, Shapiro83, Frank02}. Based on their long-term X-ray variability, BH-XRBs are broadly classified as persistent or transient sources \citep{Chen97, Zhang19}. Persistent systems exhibit relatively steady X-ray emission, whereas transient systems spend most of their lifetime in a quiescent state with low X-ray luminosities ($L_X \sim 10^{30-33}~\mathrm{erg~s^{-1}}$; \citealt{Hannikainen05, Tetarenko16}). 
In a typical outburst, transient BH-XRBs exhibit characteristic spectral and temporal evolution, tracing a `q-shaped' hysteresis pattern in the Hardness-Intensity Diagram (HID). The source initially appears in the low-hard state (LHS), followed by transitions through the hard-intermediate (HIMS) and soft-intermediate (SIMS) states before reaching the high-soft state (HSS) near the peak luminosity \citep[and references therein]{Belloni05, Done07, Nandi12, Nandi24, D13, D15, Debnath-etal2026a, D26, Sreehari19, Bhuvana25, Nandi2026}. During the decay phase, the source typically retraces the sequence in the reverse direction before returning to quiescence. However, some outbursts, known as `hard-only' or `failed' outbursts, do not reach the soft state and remain confined to the harder states \citep{Tetarenko16, D17, D23, Radhika18, Chatterjee19, Chatterjee21, Jana21, Alabarta21, Bhowmick22, Wang22}.

The temporal evolution of BH-XRBs during outbursts is characterized by rapid X-ray variability on time scales ranging from milliseconds to seconds \citep{Remillard06}. Their Fourier-transformed power density spectra (PDS) exhibit different types of noise components and Quasi-periodic Oscillations (QPOs; $\nu_{\rm QPO}\sim0.01$--30 Hz) \citep{Van89, Hasinger90}, whose properties are strongly correlated with the spectral state \citep{Remillard06, Nandi12, Nandi24, Iyer15, PMajumder26, Nath26}. Based on their timing characteristics, low-frequency QPOs (LFQPOs) are classified into type-A, -B, and -C \citep{Wijnands99, Homan01, Remillard02, Nandi2026}. Type-C QPOs are predominantly observed during LHS and HIMS and are often associated with steady, compact radio jets \citep{Fender04}. In contrast, type-B and type-A QPOs are typically detected in the SIMS \citep{Nandi12, Gao14}, with type-B QPOs frequently linked to transient relativistic jet ejections near the `jet-line' in the HID, while type-A QPOs generally precede these ejections \citep[and references therein]{Fender09, Choudhury25}. The HSS exhibits weak variability and is generally devoid of detectable LFQPOs \citep{Belloni05, Nandi12}.

The broadband X-ray spectrum of BH-XRBs provides key insights into the geometry and radiative processes of the accretion flow. It is primarily composed of a multi-temperature disk blackbody and a non-thermal power-law component. The thermal emission originates from the standard Keplerian accretion disk \citep{NT73, SS73}, while the power-law tail is produced by inverse-Compton scattering of soft disk photons by `hot' electrons in the corona (i.e. Compton cloud) \citep[and references therein]{ST80, ST85, Zdziarski85, CT95, Done07, Iyer15}. This Comptonizing region is also considered the launch site of jets and outflows \citep{C99, Radhika16, Jana17, Chatterjee19, D20, D21, Das21}. In addition, the spectrum often exhibits reflection features, including a Compton reflection hump at $\sim15$--$30$~keV \citep{Lightman88, George91, Matt91, Garcia10}, with a low energy cut-off (below $\sim10$~keV due to photoelectric absorption), and a fluorescent Fe K$\alpha$ emission line at $\sim6.4$~keV, arising from the reflection of Comptonized photons off the accretion disk \citep{Fabian89, Reynolds03, Garcia10, Garcia14}. The reflected photons may also originate from returning radiation, where the strong gravity of a rapidly spinning black hole bends photons emitted by the inner disk back onto its surface, then scattering them toward the observer \citep{Cunningham76, Agol00, Dauser22, Zhao26}.

Furthermore, X-ray polarimetry provides an additional probe of the accretion geometry in BH-XRBs. The launch of \textit{IXPE} \citep{Weisskopf22}, operating in the $2$--$8$~keV band, has enabled detailed polarization measurements of BH-XRBs across different spectral states. Most sources exhibit low polarization degrees (PDs) with Swift~J1727.8--1613 and Cyg~X--1 showing $\sim4\%$ in the LHS, decreasing to $\sim0.5\%$ and $\sim2.5\%$, respectively, in the HSS \citep{Krawczynski22, Veledina23, Ingram23, Svoboda24, Jana24, Majumder26}. Similarly, GX~339--4 and MAXI~J1744--294 show low PDs of $\sim1.3\%$ in the SIMS and HSS, while LMC~X--3 exhibits $\sim2.4\%$ in the soft state \citep{Veledina24, Jana24, Marra24, Majumder24}. In contrast, a few systems exhibit unusually high polarization, including Cyg~X--3 ($\sim21.4\%$ in SIMS), IGR~J17091--3624 ($\sim9$–$11\%$ in LHS), and 4U~1630--47 ($\sim8.3\%$ in HSS). These high polarization levels suggest diverse coronal and jet geometries, including wind-driven, slab-like, and wedge-shaped coronae \citep{Kushwaha23, Veledina24, Ewing25, D25}, as well as returning radiation around rapidly spinning black holes in the softer states \citep{Podgorny26b}.

The degree of polarization (PD) often exhibits energy dependence, with higher values observed at higher energies, particularly in the X-ray and $\gamma$-ray bands (e.g., Cyg~X-1; \citealt{Long80,Laurent11,Jourdain12,Chauvin18,Krawczynski22,Chattopadhyay24}). This dependence provides valuable insights into the geometry of the emitting region, the underlying emission mechanisms, and general relativistic effects on photon propagation \citep[][and references therein]{Chauvin18,Majumder26}. In particular, photons originating near a rapidly spinning black hole experience strong relativistic effects that can significantly modify their observed polarization properties.

Motivated by the rich timing, spectral, and polarimetric characteristics exhibited by BH-XRBs, as discussed above, we focus on the Galactic transient black hole candidate (BHC) GRS~1739--278 which serves as an excellent laboratory for investigating accretion dynamics.
The source was discovered by the \textit{SIGMA} telescope aboard \textit{Granat} during observations of the Galactic Center in 1996 \citep{Paul96, Vargas97}. Optical and infrared observations place it at a distance of $\sim8.5$~kpc \citep{Marti97}, while spectral and timing studies with \textit{RXTE} established its BHC nature \citep{Borozdin98} alongwith detection of LFQPO at $\sim 5$~Hz during $1996$ outburst \citep{Borozdin00}. The source has undergone several outbursts, including a well-studied complete outburst in $2014$--$2015$, during which the evolution of LFQPO, variable absorption ($N_{\rm H}\sim1.6$--$2.3\times10^{22}~\mathrm{cm^{-2}}$), an upper limit on the black hole mass of $18.3~M_\odot$, an inclination below $60^\circ$, and a spin parameter of $a_k=0.8\pm0.2$ were reported \citep{Borozdin00, Miller15, Yan17, Mereminskiy18, Wang18}. A transient radio jet was also detected during the decay of the 2023 `failed' outburst \citep{Hughes23}. More recently, the \textit{Einstein Probe} detected two successive outbursts in 2025 \citep{Cheng25}, prompting extensive follow-up observations with \textit{NuSTAR}, \textit{Insight-HXMT}, and \textit{IXPE}. 
Recent \textit{IXPE} observations of GRS~1739--278 measured a polarization degree (PD) of $\sim2.3\%$ and a polarization angle (PA) of $\sim62^\circ$ in the $2$--$8$ keV band \citep{Zhao26}.
Their spectro-polarimetric analysis, combining \textit{IXPE} and \textit{NuSTAR} observations with the relativistic \texttt{relxillNS} model, suggests that returning radiation may contribute to the observed polarization, offering insights into the accretion disk geometry.

In this paper, we present a comprehensive spectroscopic, timing, and polarimetric study of the Galactic transient GRS~1739--278 during its $2025$--$2026$ outburst using X-ray observations from \textit{MAXI}/GSC, \textit{Swift}/XRT, \textit{NuSTAR}, \textit{IXPE}, and \textit{Insight}-HXMT. We use \textit{MAXI}/GSC daily-averaged light curves and HID to investigate the long-term evolution of the source, comparing its hardness ratio and state transitions with those of historical outbursts. The HID reveals distinct evolutionary patterns and peak luminosity. Using \textit{Insight}-HXMT observations, we report the first detection of LFQPOs during the 2025–2026 outburst. We also perform model independent and model dependent X-ray polarization analyses using \textit{IXPE} observations from the declining phase. Finally, we investigate broadband spectral evolution across three epochs and combine the spectro-temporal-polarimetric results to propose a plausible accretion flow dynamics and radiation mechanism.

The paper is organized as follows: 
\S2 describes observation, data reduction, and analysis procedures; \S3 presents analysis results with findings; and \S4 discusses results and summarizes the conclusions.

\section{Observation and Data Analysis}

\begin{figure*}
  \centering
    \includegraphics[angle=0,width=16.0cm,keepaspectratio=true]{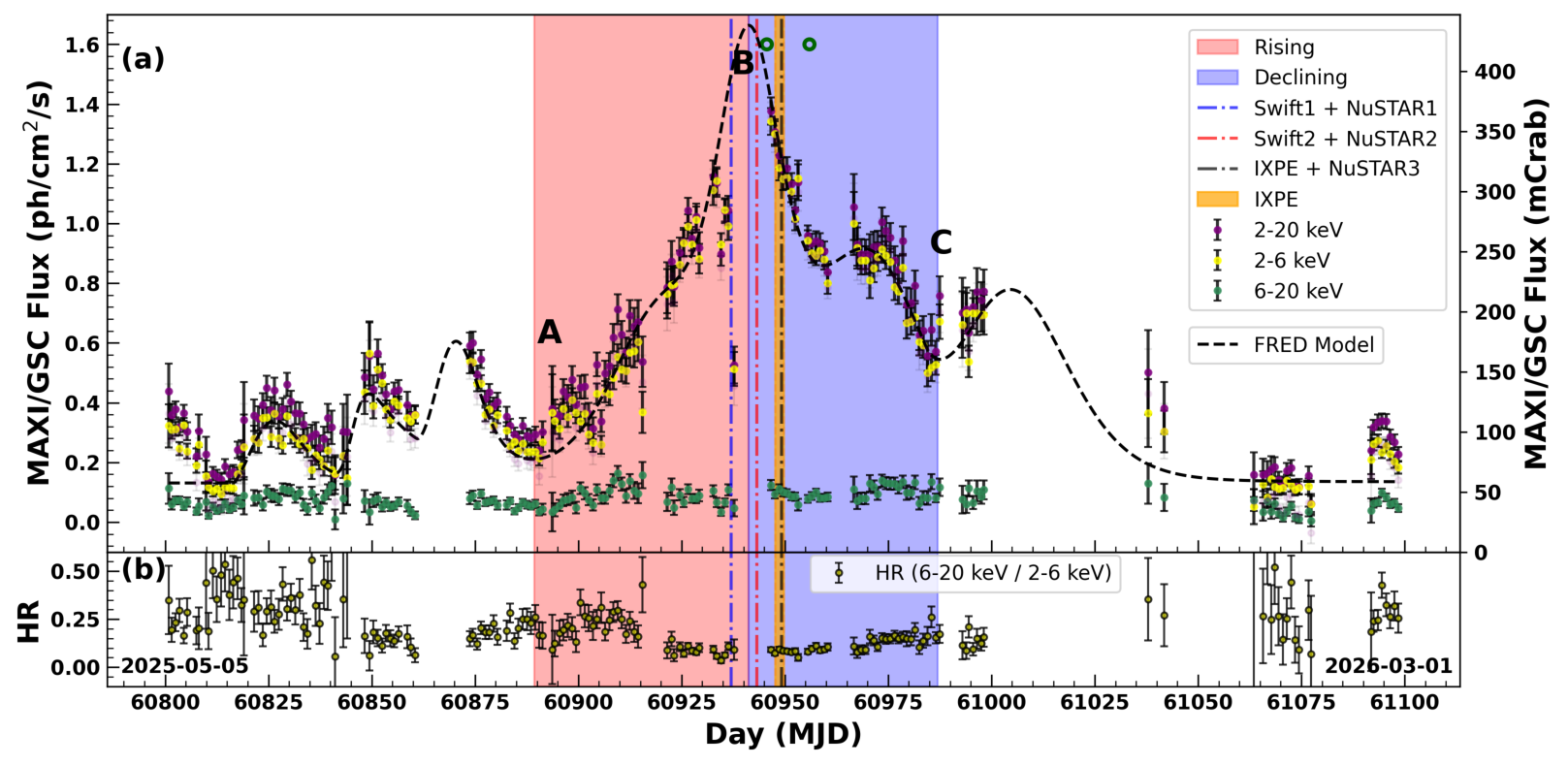}
	\caption{ Variation of (a) the daily averaged light curve of {\it MAXI}/GSC in three energy bands ($2$--$6$~keV, $6$--$20$~keV, and $2$--$20$~keV) along with the combined 7-FRED fitting profile (black dashed line)
	and (b) the hardness ratio (defined as the ratio of $6$--$20$~keV/ $2$--$6$~keV) are shown for the ongoing $2025$--$2026$ outburst 
	of GRS~$1739$--$278$. The red and blue shaded regions are considered to be the rising and declining phases respectively with A (2025-08-03), C (2025-11-08) marked as terminal points and B (2025-09-23) as the peak. The outburst lasted approximately 97 days, bounded by the onset and post-outburst minima preceding the non-detection period. The orange shaded region in the top panel marks the duration of the \textit{IXPE} observation. Three \textit{NuSTAR} observations along with two simultaneous \textit{Swift}/XRT and one \textit{IXPE} have been shown as blue, red and black dash-dot lines respectively. In the declining phase, QPOs have been detected in two \textit{Insight}-HXMT data which are shown in green circles.}
\label{fig1}
\end{figure*}

The recent $2025$--$2026$ outburst of the Galactic transient BHC GRS~1739--278 was investigated using archival observations from \textit{Swift}, \textit{NuSTAR}, \textit{IXPE}, and \textit{Insight}-HXMT. We also utilized the on-demand \textit{MAXI}/GSC data to examine the long-term evolution of the outburst (see Fig.~\ref{fig1}).

For the broadband spectral analysis, we analyzed two quasi-simultaneous \textit{Swift}/XRT and \textit{NuSTAR} observations obtained on 2025 September 18 (MJD~60936) and 2025 September 25 (MJD~60943). The \textit{Swift}/XRT exposures were approximately 17~ks and 20~ks, while the corresponding \textit{NuSTAR} observations had effective exposures of approximately 15~ks and 21~ks, respectively. We further analyzed a $\sim96$~ks \textit{IXPE} observation carried out between 2025 September 29 and October 1 (MJD~60947.63--60949.73) together with a simultaneous $\sim21$~ks \textit{NuSTAR} observation obtained on 2025 September 30 (MJD~60949). The joint spectral analysis was performed over the 1--40~keV energy range for the \textit{Swift}/XRT+\textit{NuSTAR} datasets and the 2--40~keV energy range for the \textit{IXPE}+\textit{NuSTAR} datasets.

The \textit{IXPE} observations were further utilized to investigate the spectro-polarimetric properties of the source. In addition, \textit{Insight}-HXMT monitored the source on an approximately daily cadence between 2025 September 11 (MJD~60929) and 2025 October 08 (MJD~60956). Owing to the unavailability of updated calibration files, the \textit{Insight}-HXMT data were not included in the spectral analysis and were used exclusively for timing studies.

\subsection{Data Analysis}  

In this sub-section, we present the standard data reduction procedures followed for all observations from different observatories. These include the generation of total band and energy-resolved light curves at different timebin for timing analysis as well as the extraction of energy spectra, Stokes parameters, and the corresponding response files required for spectral and polarimetric analyses. In addition, Table~\ref{table1} summarizes the simultaneous observations from different missions, including the corresponding MJDs, exposure times, energy ranges, and degree of simultaneity (DoS).

\subsubsection{MAXI/GSC}
The on-demand \textit{MAXI}/GSC light curves in the $2$--$6$~keV, $6$--$20$~keV, and $2$--$20$~keV energy bands were obtained from the MAXI-RIKEN website\footnote{\url{https://maxi.riken.jp/mxondem}}. These data were used to examine the evolution of the outburst in different energy bands, compute the HR (defined as ratio between hard to soft i.e., $6$--$20$~keV/$2$--$6$~keV fluxes), and generate the HID.

\subsubsection{Swift/XRT}

The publicly available \textit{Swift}/XRT data\footnote{\url{https://swift.gsfc.nasa.gov/analysis/}} obtained on 2025 September 18 (ObsID: 03400190001; \textit{Swift}-1) and 2025 September 25 (ObsID: 03400191001; \textit{Swift}-2) were processed using the \texttt{xrtpipeline} task available in the HEASoft software package to generate cleaned event files for windowed-timing (WT) observing modes. XRT images were extracted using the \texttt{xselect} task. For the WT-mode data, circular source and background extraction regions with radii of $30''$ and $60''$, respectively, were defined using \texttt{SAOImageDS9}\footnote{\url{https://sites.google.com/cfa.harvard.edu/saoimageds9}}. The source region was centered on the target position, while the background was extracted from a nearby source-free region. Since the source count rate remained below $\sim100$ counts s$^{-1}$, no significant pile-up correction was required. Finally, the \texttt{xselect} task was used to extract the light curves and spectra in the $0.5$--$10$~keV energy band.

\subsubsection{NuSTAR}
We analyzed three publicly available \textit{NuSTAR}\footnote{\url{https://heasarc.gsfc.nasa.gov/docs/nustar/analysis}} observations obtained on 2025 September 18 (ObsID: 91101335002; \textit{NuSTAR}-1), September 24 (ObsID: 81160303002; \textit{NuSTAR}-2), and September 30 (ObsID: 81160303004; \textit{NuSTAR}-3) using data from both FPMA and FPMB payloads. 
The data reduction was performed with the online data analysis platform SciServer using the HEASoft software package  (v6.36) provided by HEASARC.
Cleaned event files were generated using the \texttt{nupipeline} task together with the latest calibration database (CALDB20260527). Source event files were extracted from a circular region of radius $50''$ centered on the source position, while the background was extracted from an annular region with inner and outer radii of $180''$ and $240''$, respectively, using \texttt{SAOImageDS9}. Light curves with time resolutions of 10 s, 1 s, and 0.01 s along with energy spectra were extracted using the standard \texttt{nuproducts} task, which also generated the corresponding auxiliary response files (ARFs) and response matrix files (RMFs). Finally, to improve the statistical quality of the high energy spectra while maintaining a signal-to-noise ratio (SNR) of > 5, the \texttt{GRPPHA} task was used to group the spectra to a minimum of 100 counts per spectral bin. 

\begin{table}
\addtolength{\tabcolsep}{-3.5pt}
\caption{Simultaneous observations with \textit{Swift}, \textit{NuSTAR}, and \textit{IXPE}, listing the start and stop MJDs, exposure times, energy ranges, and degree of simultaneity (DoS).}

\label{table1}

\centering
\small
\vskip 0.2cm
\begin{tabular}{lcccccccc} 
\hline

Sr. & Mission& MJD$_{\text{Start}}$ & MJD$_{\text{Stop}}$ & Exp & Eng & DoS&\\ No. && &&(ks)&(keV)& ($\%$)& \\
\hline

1&\textit{Swift}-1& 60936.74 & 	60936.83 & 1.96 & 1--4 & 100 &\\
&\textit{NuSTAR}-1& 60936.71 & 	60937.14 & 15.75 & 4--40 & 12 &\\ \hline

2&\textit{Swift}-2& 	60943.03 & 60943.23 & 2.09 & 1--4 & 100 &\\
&\textit{NuSTAR}-2& 60942.94 & 	60943.51 & 21.70 & 4--40 & 9 &\\ \hline

3&\textit{IXPE}& 60947.63 & 60949.73 & 96.28 & 2--8 & 22&\\
&\textit{NuSTAR}-3& 60948.92 & 	60949.48 & 21.58 & 4--40 & 100&\\ \hline

\end{tabular}
\end{table}

\subsubsection{IXPE}

The publicly available \textit{IXPE}\footnote{\url{https://ixpeobssim.readthedocs.io}} observation (ObsID: $04251001$; \textit{IXPE}) was reduced using the latest version (v$31.0.3$) of the \texttt{ixpeobssim} software package \citep{Baldini22}. Only the Level-2 event files of Detector Units (DUs)~1 and~3 were used in the present analysis. Light curves were generated using the \texttt{xpbin} task. The source and background event files were extracted with the \texttt{xpselect} task by selecting a circular source region of radius $50''$ centered on the target position and an annular background region with inner and outer radii of $180''$ and $240''$, respectively, using \texttt{SAOImageDS9}. The \texttt{xpbin} task was further used to generate the NEFF-weighted \citep{Baldini22} Stokes spectra (\texttt{PHA1}, \texttt{PHA1Q}, and \texttt{PHA1U}) as well as the model independent \texttt{PCUBE} products for the spectro-polarimetric analysis. For spectral fitting, the \texttt{PHA1} spectra were grouped to a minimum of $50$ counts per bin using the \texttt{GRPPHA} task. The latest response files released by the \textit{IXPE} team were adopted for spectral analysis. Throughout this work, we used the newly calibrated \textit{IXPE} data (v$04$).

\subsubsection{Insight-HXMT}

We analyzed archival data from 55 \textit{Insight}-HXMT\footnote{\url{http://hxmten.ihep.ac.cn/analysis.jhtml}} observations spanning the entire $2025$--$2026$ outburst utilizing all three onboard instruments: the Low Energy (LE), Medium Energy (ME), and High Energy (HE) payloads \citep{Zhang14, Zhang20}. Light curves with timebin of 1~s and 0.01~s were extracted in the $2$--$4$, $4$--$10$, and $2$--$10$~keV energy bands for the LE instrument, the $10$--$30$~keV band for ME, and the $27$--$150$~keV band for HE. The data reduction and light curve extraction were performed using the \texttt{hpipeline} task within the HEASoft (v6.36) software package provided by HEASARC.

\section{Results}

In this Section, we present a spectro-temporal-polarimetric analysis of the source during its 2025--2026 outburst. We first examine the outburst profile and corresponding HID, comparing them with the 2014--2015 outburst. We then investigate the timing properties through PDS to identify QPOs at different outburst phases, followed by polarimetric analysis during the declining phase. Finally, we study the spectral evolution using three observations obtained near the outburst peak.

\subsection{Outburst Profile and Hardness-Intensity Diagram}

Figure~\ref{fig1}(a) shows the daily averaged \textit{MAXI}/GSC fluxes in the $2$--$20$~keV, $2$--$6$~keV, and $6$--$20$~keV energy bands from $2025$~May~$05$ (MJD~$60800$) to $2026$~March~$01$ (MJD~$61100$). Although the light curve contains several data gaps, the overall outburst profile exhibits multiple peaks with the most pronounced flux variations occurring in the softest energy band ($2$--$6$~keV). In contrast, the flux in the harder band ($6$--$20$~keV) remains comparatively stable, varying only within a narrow range throughout the outburst.
The primary outburst is divided into three phases with the boundaries marked by points A (MJD~$60890$; onset of the rise), B (MJD~$60941$; peak of the outburst), and C (MJD~$60987$; end of decay). Although point C marks the end of the outburst considered in this study, the source flux had not reached to the quiescent level observed at point A. Because of the long observational gap following this epoch, the subsequent evolution of the outburst cannot be constrained.

The peak of the outburst was determined by fitting the light curve with a fast-rise exponential-decay (FRED) model \citep{Kocevski03}. 
A combination of seven FRED profiles, accounting for pre- and post-flaring events associated with the main and subsequent outbursts along with a background component, provides a good fit ($\chi^2/dof = 200.11/134 = 1.49$). The model yields a peak at MJD~60941.2 with a maximum flux of $\sim0.4$~Crab. However, due to a gap in \textit{MAXI}/GSC coverage near the peak, the highest measured $2$–$20$~keV flux was $\sim0.3$~Crab on MJD~60947, close to the second \textit{NuSTAR} observation (see Section~\ref{sec:spectral_analysis} for details).

The red-shaded (MJD~$60890$--$60942$; $\sim52$~days) and blue-shaded (MJD~$60942$--$60987$; $\sim45$~days) regions denote the rising and declining phases of the outburst, respectively. The \textit{IXPE} observation (MJD~$60947.63$--$60949.73$; $\sim2$~days) is indicated by the orange-shaded region. The three epochs selected for the broadband spectral analysis are marked by blue (MJD~60936), red (MJD~60943), and black (MJD~60949) dash-dotted lines, corresponding to different stages of the outburst.
A significant variation of the HR (Fig.~\ref{fig1}(b)) is observed throughout the outburst, driven by the substantially larger contribution of the soft ($2$--$6$~keV) flux, which exceeds the hard ($6$--$20$~keV) flux by a factor of approximately $4$--$10$. The epochs at which QPOs were detected with \textit{Insight}-HXMT are also marked by green circles in Fig.~\ref{fig1}(a).

\begin{figure}
  \centering
    \includegraphics[angle=0,width=8.5cm,keepaspectratio=true]{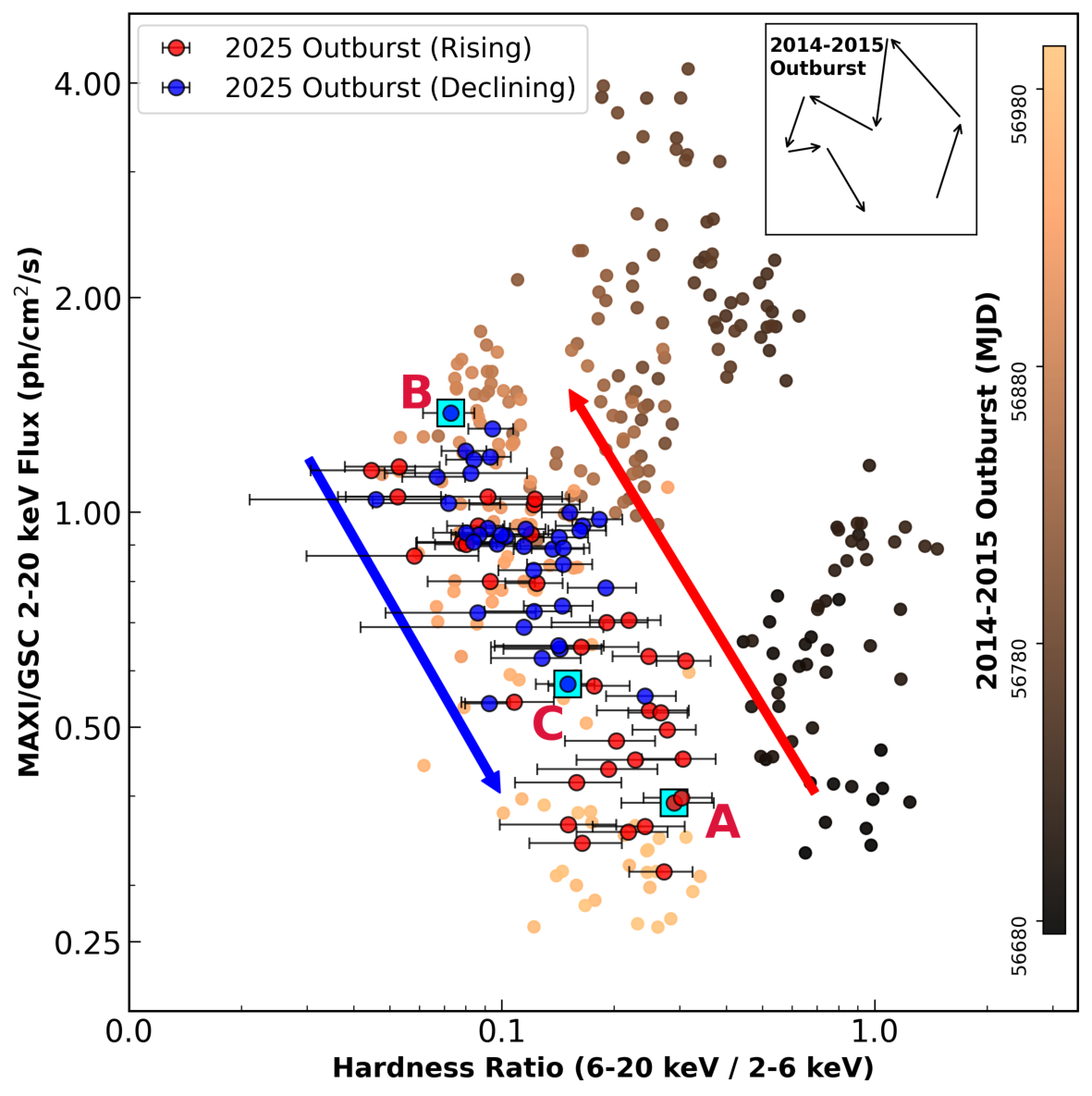}
	\caption{The HID for the 2025--2026 outburst follows a nearly diagonal track with the red and blue points representing the rising and declining phases, respectively. Points A, B, and C correspond to the same epochs marked in Fig.~\ref{fig1}. The background plot shows the HID evolution during the 2014--2015 major outburst as a function of day (MJD), while the inset highlights its canonical evolutionary track.}
\label{fig2}
\end{figure}

The HID (Fig.~\ref{fig2}) i.e., the variation of HR ($6$--$20$~keV/$2$--$6$~keV) with total flux ($2$--$20$~keV) during the $2025$--$2026$ outburst does not follow the canonical `q-shaped' track as observed in its previous major outburst in $2014$--$2015$ (background track). Transition points A, B, C are marked in aqua square boxes. The evolutionary track of the $2014$--$2015$ outburst (inset of Fig.~\ref{fig2}) shows the source transitioned between different spectral states (LHS, HIMS, SIMS, HSS) in a complete or type-I outburst of transient BH-XRB \citep[see for e.g. Figure 2 of][]{Nandi12}. 

During the 2025--2026 outburst, the source evolved from the LHS/HIMS to the SIMS during the rising phase with the flux increasing to $\sim0.3$ Crab while the HR decreased from $\sim0.20$--$0.25$ to $\sim0.04$. The minimum HR occurred several days after the flux peak, indicating continued spectral softening. During the decline, the source retraced a nearly vertical track with increasing HR as the flux decreased, remaining in the HIMS (HR $\sim0.15$ at $\sim0.1$ Crab) throughout the observations with no confirmed transition back to the LHS. Unlike the 2014--2015 outburst, which followed the canonical `q'-shaped HID (see inset and background plot of Fig.~\ref{fig2}) through the HIMS, SIMS, HSS, and back to the LHS, the 2025--2026 outburst exhibited a `diagonal' HID track with less distinct state transitions and no evidence for an HSS. A $\sim35$ days observational gap after MJD 60987 prevents confirmation of any subsequent return to the LHS (see Fig.~\ref{fig1}).

\subsection{Power Spectral Analysis}

We have analyzed the PDS of the archival \textit{Insight}-HXMT observations in the total energy band ($2$--$35$~keV) and searched for QPOs in the energy-resolved bands of $2$--$4$~keV, $4$--$10$~keV, and $10$--$30$~keV. The PDS were generated from light curves with a time resolution of 0.01~s using the \texttt{XRONOS} task \texttt{powspec} (part of the HEASoft software package) corresponding to a Nyquist frequency of 50~Hz. The PDS were computed with \texttt{norm=2} and geometrically rebinned by a factor of 1.05, yielding power spectra in units of rms$^{2}$~Hz$^{-1}$ without subtracting the white-noise level.
Among the \textit{Insight}-HXMT observations analyzed, two exhibit weak signatures of QPOs in the total energy band. These features become more prominent in the soft $2$--$4$~keV band indicating that the variability is stronger at lower energies. We also examined the PDS of the \textit{NuSTAR} observations in the total ($3$--$79$~keV) and energy-resolved ($3$--$6$, $6$--$10$, and $10$--$30$~keV) bands; however, no statistically significant QPOs were detected in any of the observations.

\begin{figure}
  \centering
    \includegraphics[angle=0,width=8.5cm,keepaspectratio=true]{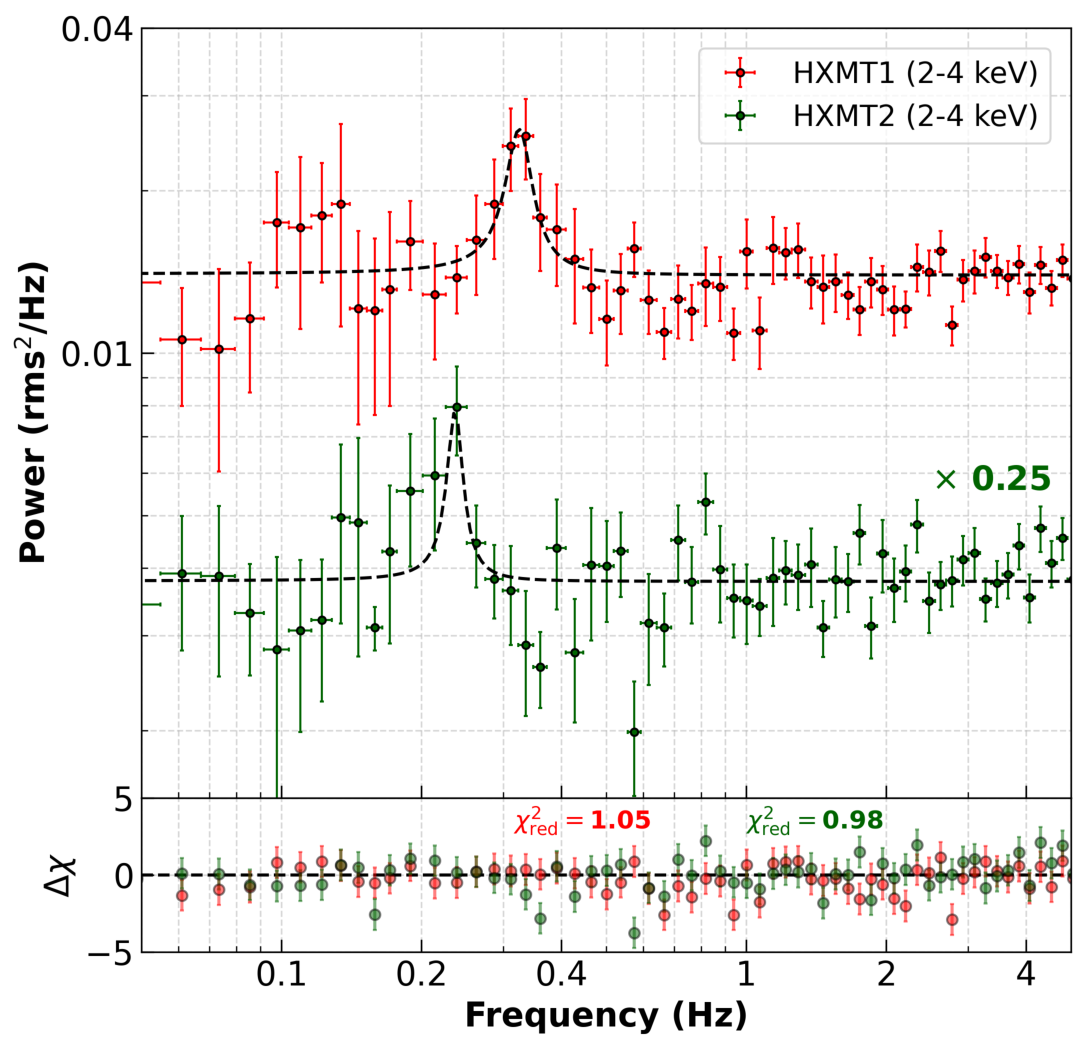}
	\caption{The upper panel shows Fourier-transformed PDS of two \textit{Insight}-HXMT observations indicating an evolution in QPO frequency. The QPO centroid frequency (in $2$--$4$~keV band) decreases from $0.32$~Hz in HXMT-1 to $0.23$~Hz in HXMT-2 in the declining phase of the outburst. The $\Delta \chi$ variation for the two observations are shown in the lower panel.}
\label{fig3}
\end{figure}

The PDS were fitted with a combination of \texttt{Lorentzian} and \texttt{Constant} models to characterize the QPO properties \citep{Van94,Belloni02}. The \texttt{Lorentzian} component is parameterized by its centroid (LC), width (LW), and normalization (LN), from which the QPO centroid frequency ($\nu_{\rm QPO}$), full width at half maximum (FWHM), and power were derived.
The best-fitting QPO parameters including the centroid frequency ($\nu_{\rm QPO}$), quality factor ($Q=\nu_{\rm QPO}/{\rm FWHM}$), fractional rms amplitude (rms$_{\rm QPO}$=$100\sqrt{\pi \times {\rm power} \times {\rm FWHM}/2}$), and detection significance ($\sigma={\rm LN}/err_{\rm neg}$), calculated following \citet{Belloni12} and \citet{Sreehari19}, are listed in Table~\ref{table2}. Figure~\ref{fig3} shows the evolution of the QPOs during the declining phase of the outburst with the centroid frequency decreasing as the outburst progresses. The best-fitting models are overplotted as black dashed lines. For clarity, the PDS of HXMT-2 (Obs ID: P081402801401) has been scaled by a factor of 0.25 relative to that of HXMT-1 (Obs ID: P081402800803). The corresponding residuals, expressed as $\chi^{2}$ variations are shown in the lower panel.

\begin{table}
\addtolength{\tabcolsep}{-5.5pt}
\centering
\caption{Model fitted QPO results.}
\small
\vskip -0.3cm
\begin{tabular}{lcccccccc} 
\hline
Serial & ObsID & UT & MJD$_{\text{Avg}}$ & Eng & $\nu_{\rm QPO}$ & Q & rms$_{\rm QPO}$ & $\sigma$ \\ 
No.& & (mm/dd) &&(keV)&(Hz)&&(\%)& \\ \hline

1 & HX0803 & 09/27 & 60945.57 & 2--4 & $0.32_{-0.01}^{+0.02}$ & 8 & 3.17 & 3.36 \\
2 & HX1401 & 10/07 & 60955.85 & 2--4 & $0.23_{-0.01}^{+0.01}$ & 11.5 & 2.96 & 2.66 \\ \hline

\end{tabular}
\vspace{0.2cm}
\small

\leftline{HX=P08140280 are pre-fixes of \textit{Insight}-HXMT ObsIDs.}
\label{table2}
\end{table}

During the 2025--2026 outburst of GRS~1739--278, low-frequency QPOs were detected only in two \textit{Insight}-HXMT observations. HXMT-1 (MJD~$60945.57$), obtained near the outburst peak, exhibits a persistent $0.32$ Hz QPO throughout the $\sim6$ ks exposure, while HXMT-2 (MJD~60955.85) shows a broader QPO at $0.23$ Hz that becomes narrower and more coherent during a $\sim0.4$ ks interval (800--1200 s; see Fig.~\ref{fig3}; Table~\ref{table2}). The QPOs are detected exclusively in the 2--4 keV band, with no significant detections above 4 keV. In the broader 2--30 keV band, they appear weaker and broader indicating that the variability is dominated by soft X-ray emission. The two QPOs have detection significances of $\sim3.4\sigma$ and $\sim2.7\sigma$, with quality factors of $\sim8$ and $\sim11.5$, respectively.

To classify the detected QPOs, we adopt the scheme of \citet{Casella04}. Although their centroid frequencies lie in the sub-Hz regime, the QPOs exhibit quality factors ($Q\gtrsim6$), fractional rms amplitudes ($\sim4\%$), and a weak red-noise continuum consistent with type-B QPOs. In contrast to type-C QPOs, QPOs are prominently detected only in the soft 2--4 keV band and are not significant in the broader energy band, where type-C QPOs are typically strongest. Given their spectral state association with the SIMS \citep{Nandi12,Choudhury25} and timing properties, we suggest that these QPOs represent a low-frequency analogue of type-B QPOs.

\begin{figure*}[!ht]
  \centering
  \begin{minipage}[c]{10cm}
    \includegraphics[angle=0, width=\textwidth, keepaspectratio=true]{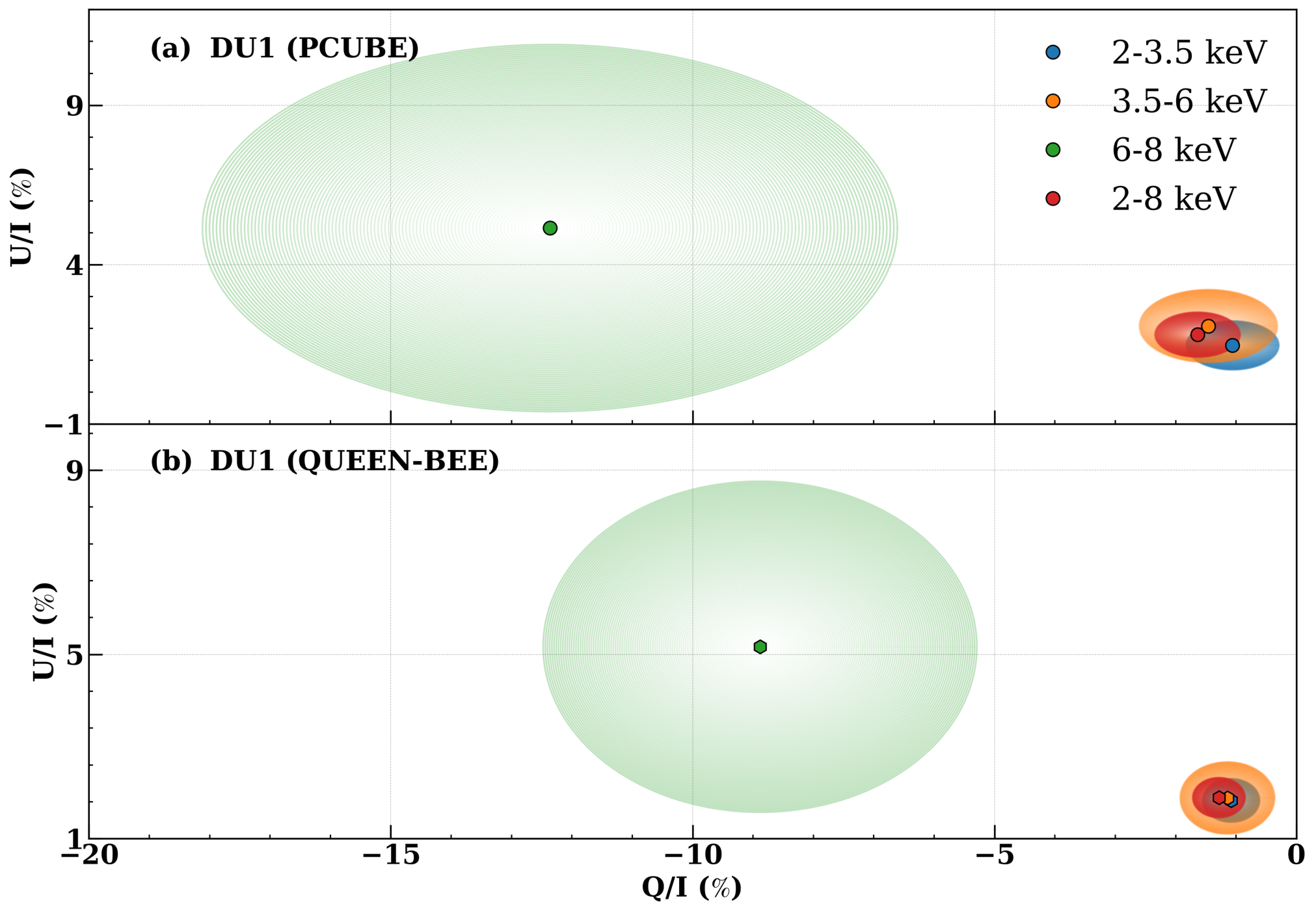}
  \end{minipage}
  \begin{minipage}[c]{6.5cm}
    \includegraphics[angle=0, width=\textwidth, keepaspectratio=true]{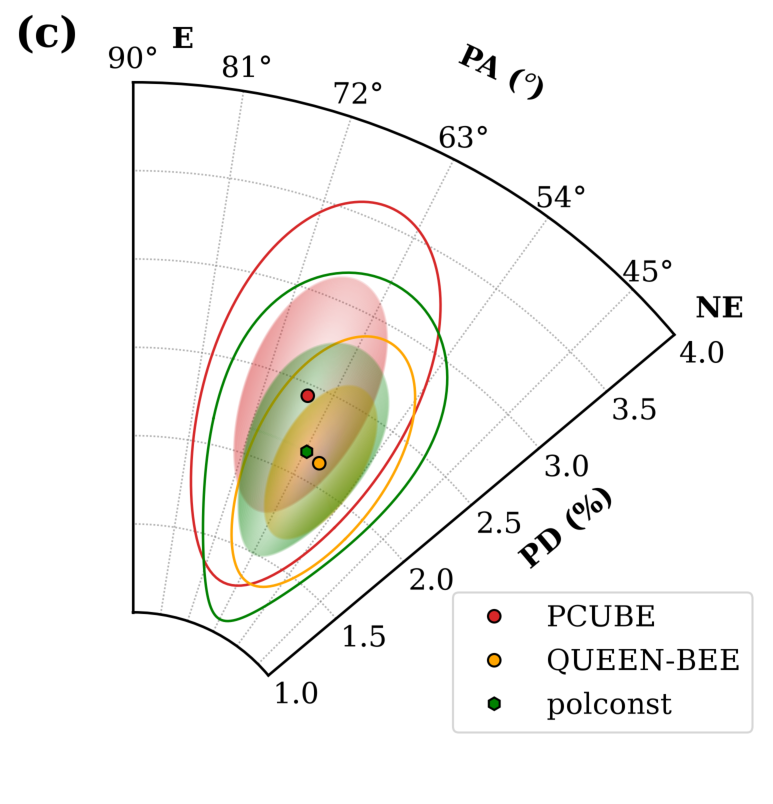}
  \end{minipage}
  
  \caption{In the \textit{left} panel, panels (a) and (b) show the (\textit{Q/I})-(\textit{U/I}) variations in the 2--3.5, 3.5--6, and 6--8 keV energy bands, with $1\sigma$ confidence levels estimated using the \texttt{PCUBE} and \texttt{QUEEN-BEE} methods, respectively. In the \textit{right} panel, panel (c) presents the polarization degree (PD) and polarization angle (PA) derived in the 2--8~keV energy band using the model independent methods (\texttt{PCUBE} and \texttt{QUEEN-BEE}) and the model dependent method (\texttt{polconst}), along with the corresponding  $1\sigma$ and $2\sigma $ confidence contours.
  }
  
  \label{fig4}
\end{figure*}

\subsection{Spectro-polarimetric Study}

We investigated the X-ray polarization properties of the source during the declining phase of the outburst using the model independent \texttt{PCUBE} algorithm in \texttt{xpbin} \citep{Baldini22} and the Bayesian nested-sampling framework \texttt{QUEEN-BEE} considering constant EVPA model \citep{Ravi25}. The analysis was performed in four energy bands: $2$–$3.5$, $3.5$–$6$, $6$–$8$, and $2$–$8$~keV. Using \texttt{PCUBE} generated polarization-cube FITS files, we derived the PD, PA, normalized Stokes parameters ($Q/I$ and $U/I$), minimum detectable polarization (MDP; 99\% confidence), and detection significance (SIGNIF). \texttt{QUEEN-BEE} uses level-2 event files and modulation response files to estimate the PD, PA, $Q/I$, and $U/I$. To avoid the ambiguity arising from the $180^\circ$ periodicity of PA, we present the energy dependence of $Q/I$ and $U/I$ obtained with both methods (\texttt{PCUBE} and \texttt{QUEEN-BEE}) in Figure~\ref{fig4}(a, b).


\begin{figure}
  \centering
    \includegraphics[angle=0, width = 8.5cm, keepaspectratio=true]{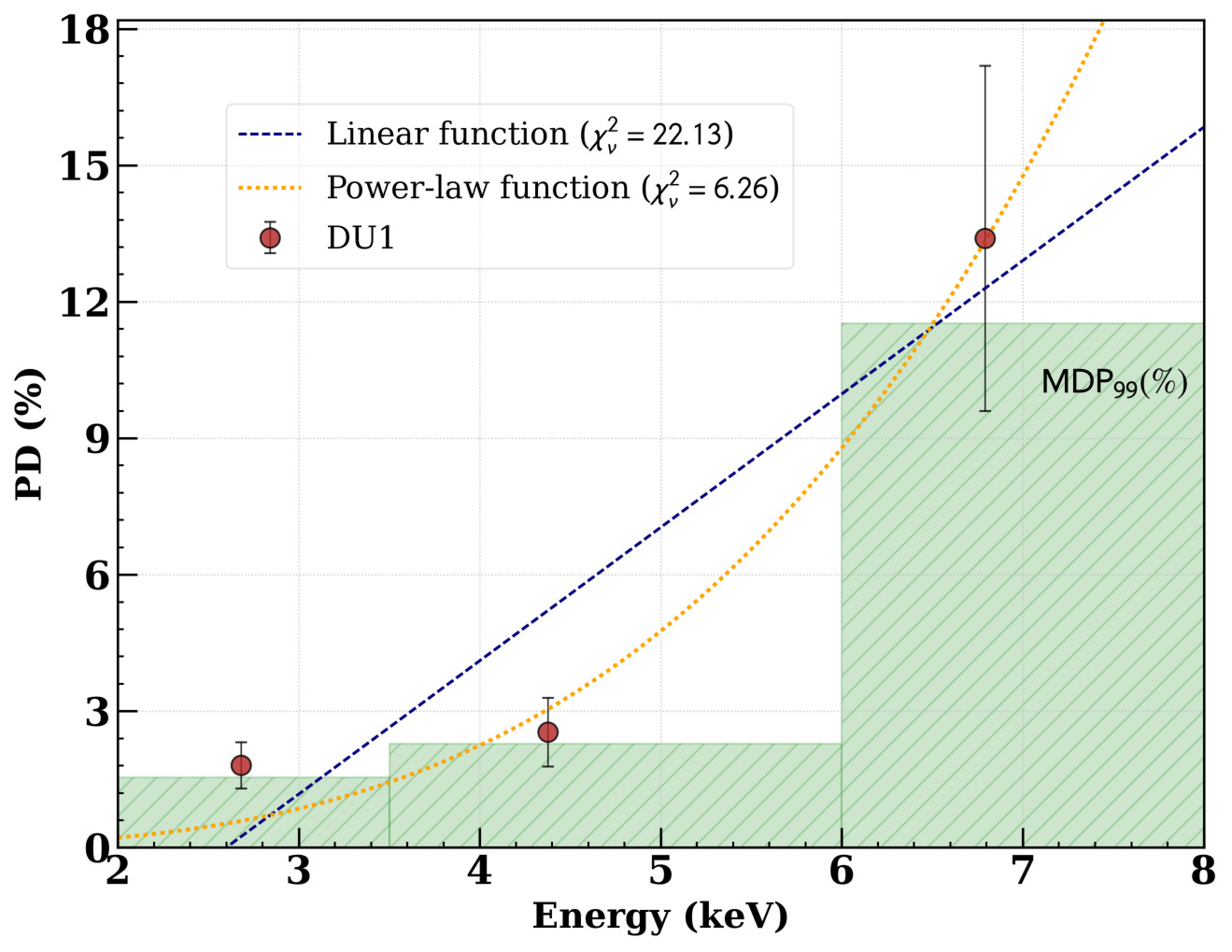}
     \caption{
     The energy dependence of the polarization degree (PD) in the 2--3.5, 3.5--6, and 6--8~keV bands is shown, as estimated using \texttt{PCUBE}. To investigate the evolution of PD with energy, a linear model and a power-law model are fitted to the data. See text for details.
     }
     
\label{fig5}
\end{figure}

The measured $Q/I$ ($U/I$) values in the $2$–$3.5$, $3.5$–$6$, $6$–$8$, and $2$–$8$~keV bands are $-1.06$ ($1.47$), $-1.46$ ($2.08$), $-12.4$ ($5.15$), and $-1.63$ ($1.81$)\%, respectively, for \texttt{PCUBE}, and $-1.08$ ($1.83$), $-1.14$ ($1.88$), $-8.88$ ($5.16$), and $-1.28$ ($1.89$)\% for \texttt{QUEEN-BEE} (see Table~\ref{table3}). Both methods reveal enhanced Stokes parameter magnitudes in the $6$–$8$~keV band, with all measurements significant at the $1\sigma$ level.
In the broad $2$–$8$~keV band, \texttt{PCUBE} yields a PD of $(2.43\pm0.47)\%$ and a PA of $(66.1\pm5.4)^\circ$, corresponding to a $\sim5\sigma$ detection. \texttt{QUEEN-BEE} yields a slightly lower PD of $(2.28\pm0.28)\%$ and a PA of $(62.0\pm3.4)^\circ$. The two estimates are consistent within $1\sigma$ and agree with the results of \citet{Zhao26}.

The energy-resolved PD and PA obtained using both model independent methods are presented in Table~\ref{table3}. The \texttt{PCUBE} results show an increase in PD from $(1.81\pm0.51)\%$ in the $2$–$3.5$~keV band to $(2.53\pm0.75)\%$ in the $3.5$–$6$~keV band and $(13.3\pm3.8)\%$ in the $6$–$8$~keV band. The corresponding PAs are $(62.9\pm8.1)^\circ$, $(62.5\pm8.5)^\circ$, and $(78.6\pm8.1)^\circ$, respectively. To characterize the energy dependence, we fitted the PD with linear ($y=mx+c$) and power-law ($y=Ax^k$) models (see Fig.~\ref{fig5}). The resulting chi-square values ($\chi^2_{lin}=22.13$ and $\chi^2_{po}=6.26$) indicate that the power-law model provides a better fit, although the energy dependence may reflect a more complex underlying radiation mechanism.

\begin{table}
\addtolength{\tabcolsep}{-4.0pt}
\centering

\caption{Polarimetric result of GRS~1739--278 in different energy bands. Model independent analysis is performed using \texttt{PCUBE} and \texttt{QUEEN-BEE} methods and model dependent analysis is performed using polconst model in XSPEC. Here, PD, $Q/I$, $U/I$, and MDP represent the polarization degree (in percent), normalized Stokes parameters, and the minimum detectable polarization at 99\% confidence, respectively. PA denotes the polarization angle (in degrees), and SIGNIF represents the polarization detection significance (in $\sigma$).}

\small
\vskip 0.5cm
\begin{tabular}{lccccc}
 \hline
Parameters        & $2$–$3.5$~keV       & $3.5$–$6$~keV        &  $6$–$8$~keV       & $2$–$8$~keV       \\
 \hline
	\multicolumn{5}{c}{Model independent result} \\
 \hline
	\multicolumn{5}{c}{\texttt{PCUBE}}\\
 \hline
PD(\%)       	  &  1.81${\pm0.51}$ &  2.53${\pm0.75}$  &  13.3${\pm3.8}$  &  2.43${\pm0.47}$ \\
PA($^\circ$)   	  &  62.9${\pm8.1}$ &  62.5${\pm8.5}$  &  78.6${\pm8.1}$  &  66.1${\pm5.4}$ \\
Q/I(\%)       	  & -1.06${\pm0.51}$ & -1.46${\pm0.75}$  & -12.4${\pm3.8}$  & -1.63${\pm0.47}$ \\
U/I(\%)       	  &  1.47${\pm0.51}$ &  2.08${\pm0.75}$  &  5.15${\pm3.8}$  &  1.81${\pm0.47}$ \\
MDP(\%)       	  &  1.54             &  2.29              &  11.5              &  1.42             \\
SIGNIF ($\sigma$) &  3.13             &  2.92              &  3.09              &  4.85             \\
 \hline
	\multicolumn{5}{c}{\texttt{QUEEN-BEE}}\\
 \hline
$\rm PD(\%)$   & 2.12${\pm0.33}$ & 2.20${\pm0.53}$&10.2${\pm2.25}$&2.28${\pm0.28}$&\\
$\rm PA(^\circ)$ & 60.2${\pm4.5}$ & 60.6${\pm6.7}$&74.9${\pm6.3}$&62.0${\pm3.4}$& \\
Q/I(\%)       	  & -1.08${\pm0.33}$ & -1.14${\pm0.51}$  & -8.88${\pm2.24}$  & -1.28${\pm0.27}$ \\
U/I(\%)       	  &  1.83${\pm0.33}$ &  1.88${\pm0.53}$  &  5.16${\pm2.27}$  &  1.89${\pm0.28}$ \\
\hline
\hline
	\multicolumn{5}{c}{Model dependent result}\\
\hline
$PD_{\rm \texttt{polconst}}(\%)$   & ---                & ---               & ---                & 2.15${\pm0.43}$   \\
$PA_{\rm \texttt{polconst}}(^{\circ})$   & ---                & ---               & ---                & 62.8${\pm6.01}$   \\
\hline
\end{tabular}
\label{table3}
\end{table}

To investigate the model dependent polarization properties of GRS~1739--278, we performed spectro-polarimetric fitting in \texttt{XSPEC} \citep{Arnaud96} using the \texttt{polconst} model. Source and background Stokes spectra ($I$, $Q$, and $U$) in the $2$–$8$~keV band were generated for \textit{IXPE} DUs 1 and 3 using the \texttt{PHA1}, \texttt{PHA1Q}, and \texttt{PHA1U} algorithms of the \texttt{xpbin} task in \texttt{ixpeobssim}.
Table~\ref{table5} summarizes the fitted parameters for four configurations: \textit{IXPE} (DU1), \textit{IXPE} (DU1+DU3), \textit{IXPE} (DU1) combined with \textit{NuSTAR} in the $4$–$10$~keV band, and the same combination over $4$–$40$~keV. For \textit{IXPE} DU1 alone, model M1: \texttt{TBabs~$\times$~ constant(polconst~$\otimes$~diskbb)}, yields $\mathrm{PD}=(2.15\pm0.43)\%$ and $\mathrm{PA}=(62.8\pm6)^\circ$, consistent with the model independent results in $2$–$8$~keV band. However, M1 does not adequately represent the spectral decomposition in the joint \textit{IXPE}+\textit{NuSTAR} fits. The fitted inner disk temperature ($T_{\rm in}$), PD, and PA remain consistent across the configurations, while the column density decreases from $N_{\rm H}\sim1.82$ for \textit{IXPE} alone to $\sim1.74$ for the joint fits. The \textit{IXPE} DU1 fit also provides a better fit than the combined DU1+DU3 analysis.
To investigate the polarization contributions of individual components within our preferred spectral decomposition, we applied a more complex model M2: \texttt{TBabs}~$\times$~\texttt{constant(thcomp}~$\otimes$~\texttt{polconst}~$\otimes$~\texttt{diskbb}~+~\texttt{polconst}~$\otimes$~\texttt{relxillNS)}. The physical parameters, including $N_{\rm H}$, black hole spin, and disk inclination, were fixed at their best-fit spectral values as obtained from M5. The direct disk component yields a PD of $\sim1.8\%$, with its lower bound pegged at zero, whereas the reflected component exhibits a substantially higher PD of $\sim14\%$. The measured PAs indicate that the reflected emission is nearly orthogonal to the direct disk emission, with $\text{PA}_{\rm dbb} = -20^\circ \pm 39^\circ$ and $\text{PA}_{\rm rel} = 66^\circ \pm 15^\circ$ ($\Delta\text{PA} \approx 86^\circ$, consistent with an orthogonal $90^\circ$ offset within errors). Imposing this orthogonality by tying the PA of reflected component to that of the disk significantly tightens the disk PA constraint to $-27^\circ\pm5.3^\circ$.

\begin{table*}[]
\addtolength{\tabcolsep}{-2.5pt}
\small
\vskip 0.2cm
\caption{Model dependent polarization results}
\label{table5}
\begin{tabular}{lccccccccc} 

\hline
\multicolumn{6}{c}{M1: \texttt{TBabs~$\times$~constant(polconst~$\otimes$~diskbb)}} \\
\hline
Mission & $N_{\rm H}$ & $T_{\rm in}$ & $\rm PD(\%)$ & $\rm PA(^\circ)$ & $\chi^2/\rm dof$ & &&&\\
\hline
\textit{IXPE} (DU1)&1.82${\pm 0.02}$ &1.00${\pm 0.01}$&2.12${\pm 0.4}$ &62.8${\pm 5.6}$& 459/442 &\\
\textit{IXPE} (DU3)&2.01${\pm 0.02}$ &0.98${\pm 0.01}$&1.74${\pm 0.1}$ &66.6${\pm 6.9}$& 469/442 &\\
\textit{IXPE} (DU1+DU3)&1.91${\pm 0.01}$ &0.99${\pm 0.01}$&1.93${\pm 0.2}$ &64.4${\pm 4.3}$& 982/888 &\\
\textit{IXPE} (DU1)+$\rm \textit{NuSTAR}^{a}$ &1.76${\pm 0.01}$ &1.02${\pm 0.01}$&2.12${\pm 0.4}$ &62.8${\pm 5.6}$& 722/590 &\\
\textit{IXPE} (DU1)+$\rm \textit{NuSTAR}^{b}$ &1.74${\pm 0.01}$ &1.02${\pm 0.01}$&2.12${\pm 0.4}$ &62.8${\pm 5.6}$& 2721/659 &\\
\textit{IXPE} (DU1)+$\rm \textit{NuSTAR}^{b}$ &1.74${\pm 0.01}$ &1.02${\pm 0.01}$&2.12${\pm 0.4}$ &62.8${\pm 5.6}$& 2721/659 &\\

\hline
\hline
\multicolumn{10}{c}{M2: \texttt{TBabs~$\times$~constant(thcomp~$\otimes$~polconst~$\otimes$~diskbb~+~polconst~$\otimes$~relxillNS)}} \\

\hline
Mission & $\rm PD_{\rm dbb}(\%)$ & $\rm PA_{\rm dbb}(^\circ)$ & $\rm PD_{\rm rel}(\%)$ & $\rm PA_{\rm rel}(^\circ)$ & $\chi^2/\rm dof$ & \\
\hline

\textit{IXPE} (DU1)+$\rm \textit{NuSTAR}^{b}$ &1.8${\pm 2}$ &-19.4${\pm 39}$& 14.2${\pm 7}$ & 65.8${\pm 15.3}$ & 763/653
&\\
\textit{IXPE} (DU1)+$\rm \textit{NuSTAR}^{b}$ &1.7${\pm 2}$ &-26.8${\pm 5.3}$& 14.2${\pm 7}$ & $=\psi_{\rm dbb}+90^\circ$ & 764/654
&\\

\hline
\end{tabular}

\vskip 0.2cm
\leftline{a = \textit{NuSTAR} energy range 4--10~keV; b = \textit{NuSTAR} energy range 4--40~keV}

\end{table*}

\begin{figure*}
  \centering
    \includegraphics[angle=0,width=8.5cm,keepaspectratio=true]{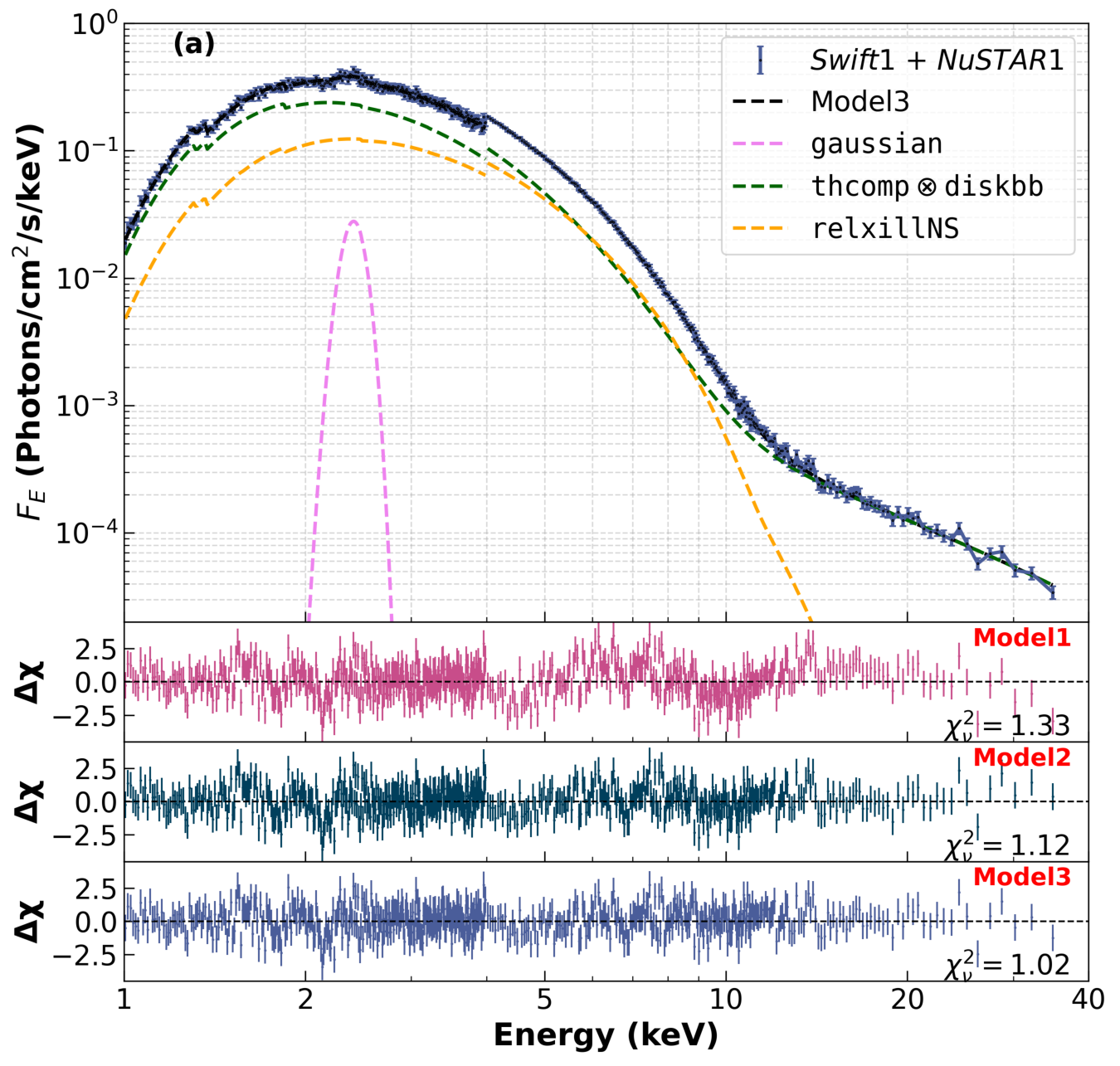}
    \includegraphics[angle=0,width=8.5cm,keepaspectratio=true]{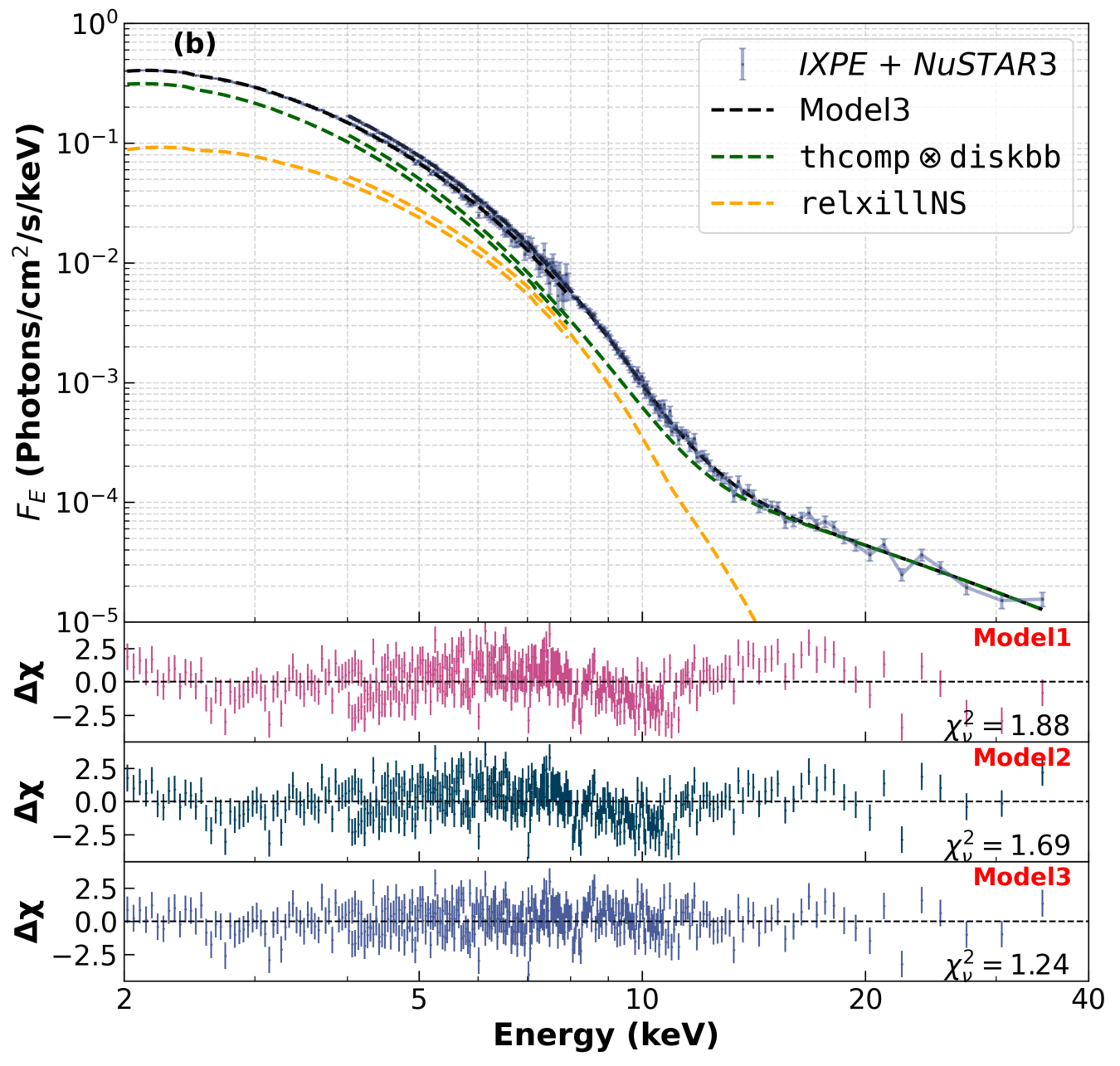}
        \caption{ (a) Combined \textit{Swift}/XRT--\textit{NuSTAR} spectra (1--40 keV) fitted with three models: M3 (\texttt{TBabs$\times$edge$\times$constant(thcomp$\otimes$diskbb + gaussian)}), M4 (\texttt{TBabs $\times$ edge $\times$ constant(thcomp $\otimes$ diskbb + gaussian + relxill)}), and M5 (\texttt{TBabs $\times$ edge $\times$ constant(thcomp $\otimes$ diskbb + gaussian + relxillNS)}). The top panel shows the unfolded spectrum of M5 fit and its spectral components for the first \textit{NuSTAR} observation (ID: NU91101335002), while the bottom panel shows the corresponding $\Delta\chi$ residuals for all models. (b) Combined \textit{IXPE}--\textit{NuSTAR} spectra (2--40 keV) fitted with the same models, excluding the \texttt{edge} and \texttt{gaussian} components. The top panel shows the unfolded spectrum of M5 fit for the third \textit{NuSTAR} observation (ID: NU81160303004) and the bottom panel displays the corresponding $\Delta\chi$ residuals.}
\label{fig6}
\end{figure*}

\subsection{Spectral Analysis}
\label{sec:spectral_analysis}

We performed a broadband spectral analysis of the source over the $1$--$40$~keV energy range by jointly fitting the \textit{Swift}/XRT and \textit{NuSTAR}/FPMA, FPMB spectra for the first two observations (MJD~60936.93 and MJD~60943.23). For the third observation (MJD~60949.20), we carried out a joint spectral analysis of the \textit{IXPE} and \textit{NuSTAR}/FPMA, FPMB data over the $2$--$40$~keV energy range. The upper energy limit of the \textit{NuSTAR} data was restricted to $40$~keV because of the increasing background contribution at higher energies. In addition, the \textit{NuSTAR}/FPMB data below $4$~keV were excluded owing to known calibration issues\footnote{\url{https://heasarc.gsfc.nasa.gov/docs/nustar/nustar_faq.html}}.

For the \textit{Swift}/XRT spectra, we considered only the $1$--$4$~keV energy range because of the low SNR ($<3$) and increasing background contribution at higher energies. A systematic uncertainty of $1\%$ was applied to all \textit{Swift}/XRT spectral fits. The feature near 1.4~keV is treated as an empirical calibration correction, likely associated with the potential residual of the Al K$\alpha$ fluorescence absorption line \citep{Godet09}. To account for this feature, an additional \texttt{edge} component was included in the spectral model. We also added a \texttt{gaussian} component centered at $\sim2.4$~keV to model residual instrumental calibration features \citep{Beardmore18} in the CCD detector associated with the Si K-edge ($1.84$~keV) and Au M-edge ($2.2$--$3.0$~keV)\footnote{\url{https://www.swift.ac.uk/analysis/xrt/digest_cal.php}} of \textit{Swift}/XRT. These additional components were not required for the third observation (\textit{IXPE} + \textit{NuSTAR}-3).

\begin{table*}[]
\addtolength{\tabcolsep}{-3.0pt} 
\vskip -0.0cm
\centering
\caption{Spectral fitted results with three models (i.e., M3, M4 and M5).}
\label{table4}
\addtolength{\tabcolsep}{-0.5pt}
\small
\vskip -0.3cm
\renewcommand{\arraystretch}{0.9}  
\begin{tabular}{lcccccccccccccccc} 
\hline
\hline
\multicolumn{3}{c}{}& \multicolumn{5}{c}{M3: \texttt{thcomp $\otimes$ diskbb }} \\
\hline
Serial No.& ObsID & MM/DD &  MJD$_{Avg}$ & Exp& $N_{\rm H}$ & $T_{\rm in}$ & $\Gamma$ & $\chi^2/$dof & &&& \\
\hline
 $1^{*}$& 91101335002& 09/18& 60936.93 & 15803 &1.90${\pm 0.02}$ &1.02${\pm 0.01}$&1.71${\pm 0.02}$ & 1042/782  &  \\
 $2^{*}$& 81160303002& 09/24& 60943.23 & 21669 &1.95${\pm 0.02}$ &1.01${\pm 0.01}$&1.75${\pm 0.02}$ & 1034/856  &  \\
 3& 81160303004& 09/30& 60949.20 & 21696 &1.85${\pm 0.01}$ &1.00${\pm 0.01}$&1.60${\pm 0.03}$ & 1363/724 &  \\
\hline
\multicolumn{3}{c}{}& \multicolumn{5}{c}{M4: \texttt{thcomp $\otimes$ diskbb + relxill}} \\
\hline
Serial No.& $N_{\rm H}$ & $T_{\rm in}$ & $\Gamma$ & $a_{\rm k}$  & $i$ & $\log\xi$ & $\chi^2$/dof & $F^{}_{\rm dbb}$ & $F^{}_{\rm nth}$ & $F^{}_{\rm ref}$ & $F^{}_{\rm bol}$ &\\
\hline
 $1^{*}$& 1.90${\pm 0.01}$ &1.02${\pm 0.02}$&1.81${\pm 0.15}$& 0.852${\pm 0.115}$& 50.8$^{\circ}{\pm 5.7}$& 3.06${\pm 0.13}$ & 873/778  &  8.56 & 0.001 & 0.22 &8.7\\
 $2^{*}$& 1.95${\pm 0.01}$ &1.01${\pm 0.01}$&1.98${\pm 0.07}$& 0.960${\pm 0.025}$& 62.8$^{\circ}{\pm 2.9}$ & 2.69${\pm 0.21}$ & 891/852  &  10.3 & 0.040 & 0.26 & 10.6\\
 3& 1.85${\pm 0.01}$ &1.00${\pm 0.01}$&1.88${\pm 0.20}$& 0.895${\pm 0.105}$& 57.0$^{\circ}{\pm 6.3}$ & 2.70${\pm 0.50}$& 1219/720 & 8.76 & 0.001 & 0.07 & 8.8\\
\hline

\multicolumn{3}{c}{} & \multicolumn{5}{c}{M5: \texttt{thcomp $\otimes$ diskbb + relxillNS}} \\
\hline
Serial No.& $N_{\rm H}$ & $T_{\rm in}$ & $\Gamma$ & $a_{\rm k}$  & $i$ & $\log\xi$ & $\log N$ & $\chi^2/$dof & $F^{}_{\rm dbb}$ & $F^{}_{\rm nth}$ & $F^{}_{\rm ref}$ & $F^{}_{\rm bol}$ &\\
\hline
 $1^{*}$& 1.80${\pm 0.09}$ &0.93${\pm 0.04}$&1.93${\pm 0.03}$& 0.984${\pm 0.052}$& 63.0$^{\circ}{\pm 4.0}$& 3.27${\pm 0.43}$ & 19.0${\pm 4.9}$ & 798/777  & 4.32 & 0.19 & 3.79 & 8.3 & \\
 $2^{*}$& 1.75${\pm 0.07}$ & 0.93${\pm 0.03}$&1.89${\pm 0.02}$& 0.950${\pm 0.050}$& 59.9$^{\circ}{\pm 6.3}$& 3.75${\pm 0.24}$ & 19.0${\pm 3.0}$ & 782/851  & 4.74 & 0.27 & 4.60 & 9.6 & \\ 3& 1.78${\pm 0.02}$ &0.94${\pm 0.02}$&2.00${\pm 0.05}$& 0.953${\pm 0.038}$& 58.5$^{\circ}{\pm 3.3}$& 2.84${\pm 0.22}$ & 19.0${\pm 3.6}$ & 898/719  & 6.54 & 0.07 & 2.07 & 8.6 &\\
\hline
\end{tabular}

\leftline{Fixed parameter values of \texttt{thcomp}: $kT_e = 20$~keV; \texttt{relxill} and  \texttt{relxillNS}: $Index1$ = $Index2$ = 6, $R_{\rm br}=15$, $R_{\rm in}=-1$, }
\leftline{$R_{\rm out}=1000$ and $A_{\rm Fe}=1.5$ are used.} 
\leftline{Units of $N_{\rm H}$ in $10^{22}$~${\rm cm}^{-2}$; $T_{\rm in}$ in keV; $R_{\rm br}$ in ISCO radius; $R_{\rm out}$ in Gravitational radius $r_{\rm g}=GM/c^2$.}
\leftline{Thermal \texttt{diskbb}, non-thermal \texttt{thcomp}, reflected \texttt{relxill} and \texttt{relxillNS} and bolometric fluxes: $F_{\rm dbb}$, $F_{\rm nth}$, $F_{\rm ref}$ and $F_{\rm bol}$}
\leftline{are in units of $10^{-9}~{\rm erg}~{\rm cm}^{-2}~{\rm s}^{-1}$ and calculated in extend energy range of $1$--$100$~keV.} 
\leftline{Note, model fitted parameter errors are the average values of the $1\sigma$ $\pm$ errors. }
\leftline{* \texttt{edge} \& \texttt{gaussian} is added to compensate \textit{Swift}/XRT instrumental features.}

\end{table*}

The combined spectra from the first two observations were fitted using three spectral models: M3: \texttt{ TBabs $\times$ edge $\times$ constant(thcomp $\otimes$ diskbb + gaussian)}, M4: \texttt{TBabs $\times$ edge $\times$ constant(thcomp $\otimes$ diskbb + gaussian + relxill)}, and M5: \texttt{ TBabs $\times$ edge $\times$ constant(thcomp $\otimes$ diskbb + gaussian + relxillNS)} (see Table~\ref{table4} and Fig.~\ref{fig6}(a)). For the third observation, the same models were employed after removing the \texttt{edge} and \texttt{gaussian} components, resulting in modified versions of M3, M4, and M5 (see Fig.~\ref{fig6}(b)).
The thermal Comptonization model (\texttt{thcomp}) was used to characterize the `hot' corona responsible for up-scattering seed photons. The reflection-only models \texttt{relxill} and \texttt{relxillNS} were included with the illuminating continuum provided by the Comptonized component and thermal blackbody emission, respectively \citep{Garcia22}. For \texttt{relxill}, the high-energy cutoff was fixed at $E_{\rm cut}=3\times kT_e$ as expected for thermal Comptonization \citep{Petrucci01}. Using the best-fit M4 and M5 models ($\chi_\nu^2\approx1$), we estimated the bolometric and component fluxes over 1--100 keV with the \texttt{cflux} convolution model in \texttt{XSPEC}.

For model M3, the hydrogen column density ($N_{\rm H}$) derived from \texttt{TBabs} is $(1.85$--$1.95)\times10^{22}$ cm$^{-2}$. The coronal electron temperature was unconstrained above 20 keV and was therefore fixed at this value. The fits yield an inner disk temperature of $T_{\rm in}\approx1.0$ keV and photon index $\Gamma=1.60$--$1.71$ across the three observations. The low covering fraction from \texttt{thcomp} indicates a weak Comptonized contribution suggesting a disk-dominated spectrum. However, residuals around $\sim15$ keV reveal a reflection component, motivating the inclusion of relativistic reflection models in M4 and M5.

In model M4, the \texttt{relxill} component was added as a reflection-only component, significantly improving the fit over M3 while keeping $N_{\rm H}$ and $T_{\rm in}$ consistent across all observations (see Table~\ref{table4}). The component fluxes show that the spectrum is dominated by thermal disk emission ($F_{\rm dbb}$) with a minimal contribution from Comptonization ($F_{\rm nth}$). However, since \texttt{relxill} assumes illumination by a non-thermal coronal continuum, it may not be physically appropriate for this disk-dominated state. Therefore, in model M5, we replaced \texttt{relxill} with \texttt{relxillNS}, which models reflection illuminated by thermal disk emission.

In the final model (M5), we replaced \texttt{relxill} with \texttt{relxillNS} linking the blackbody temperature ($kT_{\rm bb}$) to the \texttt{diskbb} inner disk temperature to model reflection illuminated by thermal disk photons. The best-fit parameters are well constrained with $N_{\rm H}\approx(1.75$--$1.80)\times10^{22}$ cm$^{-2}$, $T_{\rm in}\approx0.93$ keV, and $\Gamma=1.93$--$2.00$, consistent across all observations. The model yields a rapidly rotating black hole ($a_{\rm k}=0.95$--$0.98$), a disk inclination of ($58.5$--$63.0$)$^\circ$, and a highly ionized disk ($\log\xi=2.84$--$3.75$), consistent with previous studies \citep{Miller15, Furst16, Liu23, Zhao26}.
The derived fluxes confirm that thermal disk emission dominates the spectrum, while the Comptonized component remains weak (see Table~\ref{table4}). The bolometric flux ($F_{\rm bol}$) peaks during the second observation, increasing by $\sim15\%$ before declining by $\sim10\%$ in the third. The reflected and Comptonized fluxes show similar trends, whereas the disk flux increases steadily by $\sim10\%$ and $\sim30\%$ over the three observations.

\section{Discussion and Concluding Remarks}

In this paper, we present a comprehensive spectral, timing, and polarimetric study of the Galactic transient black hole candidate GRS~1739--278 during its $2025$--$2026$ outburst. The outburst was extensively monitored over a broadband energy range by \textit{MAXI}/GSC, \textit{Swift}/XRT, \textit{NuSTAR}, \textit{IXPE}, and \textit{Insight}-HXMT. Since its discovery in 1996, GRS~1739--278 has undergone two major outbursts (1996 and 2014--2015) along with two `failed' outbursts (2015--2016 and 2023). The $2025$--$2026$ outburst reached a peak intensity of approximately $0.3$ Crab, making it substantially fainter than its previous major outbursts. However, there is a observational data gap ($\sim11$~days) close to the peak. A multi-FRED profile is modeled for the complete $2025$--$2026$ outburst to identify the peak flux on MJD~60941.2, approximately six days before the highest measured flux was on MJD~60947. The fitting parameters also reflectes a total outburst period of $\sim97$~days in which the main and its subsequent event exhibited $\sim52$ days of rising and $\sim45$ days of declining period followed by an observational data gap.

The daily-averaged \textit{MAXI}/GSC on-demand light curves in the soft ($2$--$6$~keV), hard ($6$--$20$~keV) and total ($2$--$20$~keV) energy bands were used to investigate the evolution of the flux and HR during the outburst. 
Throughout this period, the soft X-ray flux consistently dominated over the hard X-ray flux, while the hard-band emission remained nearly constant. This behavior indicates that the $2025$--$2026$ outburst was primarily driven by enhanced thermal emission from the Keplerian accretion disk with only a minimal contribution from the non-thermal coronal component.
Accordingly, HID traces an almost `unconventional' slanted straight-line track (Fig.~\ref{fig2}), rather than the canonical `q'-shaped evolution observed during the 2014--2015 major outburst. Such behavior is characteristic of a `failed' (type-II) outburst. Unlike a canonical (type-I) outburst, which progresses through all four spectral states (LHS, HIMS, SIMS, and HSS), `failed' outbursts do not reach the HSS, and in some cases the intermediate states are also absent \citep[and references therein]{D17,D23, Chatterjee19, Chatterjee21, Jana21, Bhowmick22, Adegoke25}. This behavior is generally attributed to an insufficient supply of high-viscosity Keplerian matter from the companion star, preventing efficient cooling of the `hot' corona and thereby inhibiting the transition to the softer spectral states. Since GRS~1739--278 never reaches the HSS during its 2025 activity, we classify this event as a `failed' outburst.

We report the first detection of two `low-frequency' type-B QPOs in GRS~1739--278, observed in the 2--4 keV band during two \textit{Insight}-HXMT observations obtained near the peak and during the decay of the 2025--2026 outburst. Type-B QPOs are typically associated with the SIMS and are widely regarded as signatures of spectral state transitions \citep{Fender04, Jana21}. They have previously been detected at frequencies of $\sim2$--$3$ Hz in XTE~J1752$-$223, $\sim4$ Hz in MAXI~J1348--630 and Swift~J1658.2$-$4242, and $\sim5.4$ Hz in Swift~J1728.9$-$3613 \citep[and references therein]{Chatterjee20, Zhang21, Saha23, Choudhury25}. Sub-Hz type-B QPOs ($\sim0.9$ Hz) have also been reported in GX~339--4 during its 2002 and 2010 outbursts \citep{Motta11}. 
The origin of type-B QPOs remains uncertain and has been attributed to mechanisms such as weak coronal resonance or violations of the Rankine--Hugoniot shock conditions \citep{C15}. In GX~339$-$4, they are associated with flux enhancements and hard X-ray emission, suggesting a connection with transient jet activity \citep{Motta11}. In contrast, the type-B QPOs detected in GRS~1739--278 are detected exclusively to the soft 2--4 keV band and are absent at higher energies. Combined with the negligible contribution of the non-thermal Comptonized component (Table~\ref{table4}), this suggests that they are unlikely to originate in the `hot' corona. Instead, they may arise from magnetic pressure enhancements near the inner accretion disk during the SIMS, where repeated magnetic reconnection and toroidal flux-tube ejections disrupt the inner disk, suppress the non-thermal emission, and produce low-energy type-B QPOs \citep{Nandi01,Radhika14}. Unfortunately, the lack of contemporaneous radio observations during the 2025--2026 outburst prevents testing a possible jet connection.

Broadband spectral analysis of the three observations obtained around the peak of the $2025$--$2026$ outburst indicates that GRS~1739--278 remained in the SIMS with the thermal disk emission dominating the X-ray continuum and only a weak contribution from the Comptonized coronal component. Previous studies of the 2014--2015 outburst reported strong reflection features during the LHS and HIMS, where the reflection was attributed to illumination by non-thermal coronal photons \citep{Mereminskiy18, Liu23}. In contrast, although the \texttt{relxill} model (M4) successfully reproduces the observed reflection spectrum, it requires only a negligible Comptonized flux ($F_{\rm nth}$) while still yielding a significant reflected flux ($F_{\rm ref}$). This suggests that the observed reflection is unlikely to be reprocessed by the coronal emission.
A more physically consistent description is provided by the \texttt{relxillNS} model (M5), in which the reflector is illuminated by thermal disk photons. The combination of a relatively cool corona and a high-density accretion disk can naturally produce an excess of soft photons that drive the reflection spectrum \citep{Garcia16}. For highly spinning BH in softer spectral state, returning disk photons reflected back from the disk as well \citep{Cunningham76, Garcia22, Zhao26}. 
The evolution of the spectral fluxes further supports this picture. The reflected flux ($F_{\rm ref}$) follows the same trend as the bolometric flux ($F_{\rm bol}$) increasing toward the outburst peak before declining, whereas the thermal disk flux ($F_{\rm dbb}$) increases monotonically across the three observations. The temporary enhancement of $F_{\rm bol}$ near the peak together with the highest measured ionization parameter ($\log\xi$) may indicate the onset of a state transition. The best-fitting model also implies a rapidly spinning black hole with $a_{\rm k}\approx0.95$--$0.98$, consistent with previous measurements \citep{Miller15, Zhao26}.

The \textit{IXPE} polarimetric observations obtained during the declining phase, close to the outburst peak, yield (using \texttt{PCUBE}) a PD of $2.43\%$ at a PA of $66.1^\circ$ in the $2$--$8$~keV band, detected at a significance of approximately $5\sigma$. A clear energy dependence is observed in PD (Fig.~\ref{fig5}; Table~\ref{table3}) with a maximum of $13.3\%$ in the $6$--$8$~keV band (\texttt{PCUBE}).
Conversely, the PA variation is mostly stable for $2$--$3.5$ ($\sim 62.9^\circ$)  and $3.5$--$6$~keV ($\sim 62.5^\circ$) with a sudden change in $6$--$8$~keV band ($\sim 78.6^\circ$).  
Simultaneous broadband spectroscopy confirms that the source was in the SIMS during this observation with significant contribution from the reflected photons (see Table~\ref{table4}). 
Alongside, the spectro-polametric study taking reflection scenario into account in M2 (Table~\ref{table5}) give a more comprehensive view where the low energy regime mostly dominated by the disk photons and the returning radiation close to BH produces high $\text{PD}$ ($>10\%$) with an orthogonal $\text{PA}$ relative to direct disk photons \citep{Schnittman09,Connors20}. The combined superposition of direct disk photons and orthogonal returning radiation effectively accounts for the observed net PD and its energy-dependent rotation in the 6–8 keV band. 
The low value of PD is consistent with measurements from other black hole X-ray binaries in the SIMS, such as GX~339$-$4 ($\sim1\%$; \citealt{Mastroserio25}) and Swift~J1727.8$-$1613 ($\sim3.3\%$; \citealt{Ingram24}). Such weak polarization has been interpreted as arising from the coexistence of a radially extended, weak compact corona and a dominant accretion disk \citep{Majumder26} or from a wedge-shaped corona covering the inner disk \citep{Poutanen23}. In a planar slab corona, repeated Compton scatterings preferentially polarize outgoing photons parallel to the disk plane, reproducing the observed rising PD with energy. While standard static wedge-shaped coronal models predict only $\text{PD} \sim 1\%$ at low inclinations ($i \lesssim 30^\circ$; \citealt{Krawczynski22}), higher polarization values ($\text{PD} \sim 3\% - 6\%$) are naturally produced in extended wedge geometries if the outflowing coronal plasma is mildly relativistic ($v/c \sim 0.2 - 0.4$; \citealt{Beloborodov98}) or observed at intermediate inclinations ($30^\circ \lesssim i \lesssim 60^\circ$; J. \citealt{Poutanen23}). Multiple Thomson scatterings in the optically thick inner disk can further reduce the net polarization \citep{ST85}.
An important result is the pronounced increase in PD toward higher energies. Similar behavior in GX~339$-$4 has been attributed to coronal geometry \citep{Mastroserio25}, while the observed net polarization is generally interpreted as a superposition of orthogonally polarized disk and coronal emission \citep{Ingram24}. However, our spectral analysis indicates that the Comptonized component contributes only $\sim1$–$3\%$ of the total flux (see Table~\ref{table4}), making it less likely to be the primary origin of the high-energy polarization. Instead, the increasing PD is more plausibly associated with the substantial contribution of the thermal reflection component inferred from the spectral fits. Although wind-driven scattering has been invoked to explain the high soft-state polarization observed in sources such as 4U~1630–47 \citep{Kushwaha23}, our results favor a combination of returning radiation and scattering in a wedge-shaped corona as the more plausible explanation for the polarization behavior of this source.


\begin{figure*}[!ht]
  \centering
    \includegraphics[angle=0, width = 16.0cm, keepaspectratio=true]{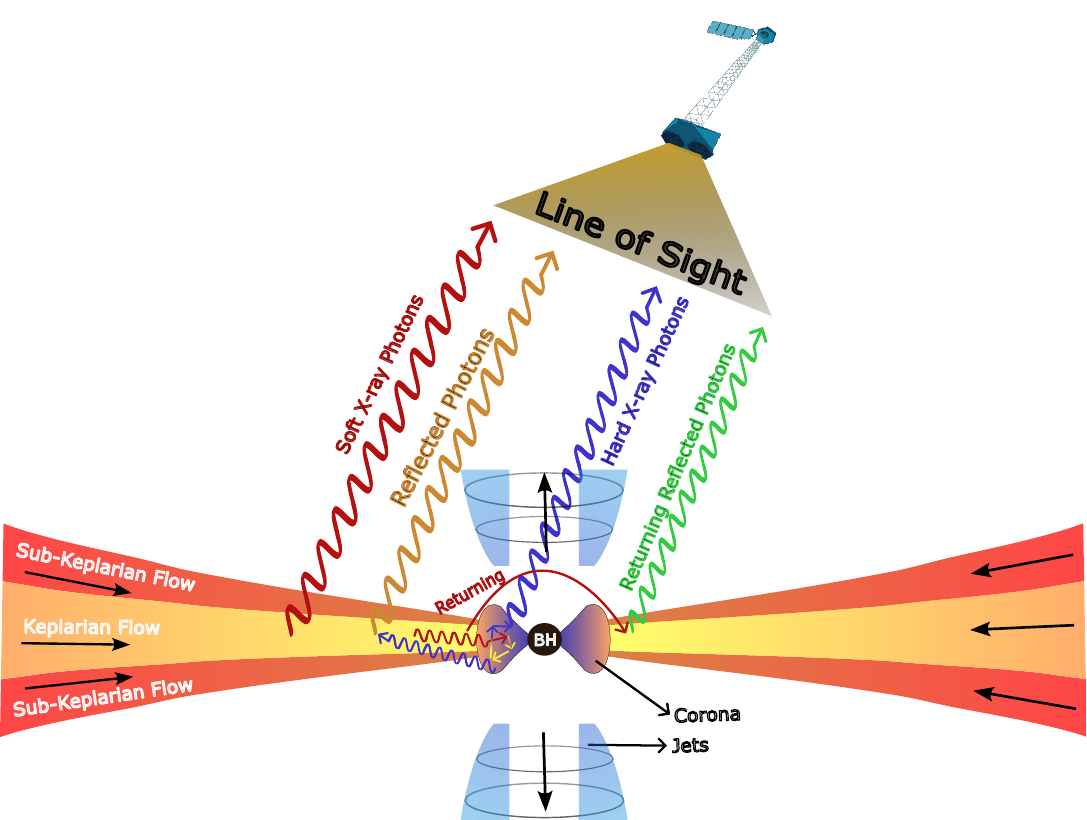}
     \caption{Schematic representation of a possible disc-jet-corona configuration for GRS~1739--278 within two-component advective flow framework. The Keplerian and sub-Keplerian flow components, the corona, and the possible jets are shown in different colors. Arrows indicate the origin and propagation of soft (red), hard (blue) and reflected (orange), and returning reflected (green) X-ray photons. The two-component accretion flow geometry is adapted from \citet{CT95}.}
\label{fig7}
\end{figure*}

Combining the results of our spectro-temporal analysis, we interpret the accretion dynamics of GRS~1739--278 within the two-component advective flow (TCAF) framework \citep{CT95}. Based on hydrodynamic solutions considering radiative transfer equations, this model (see Fig.~\ref{fig7}) describes how the interplay between the Keplerian and sub-Keplerian flows can physically produce the Comptonizing corona and drive its oscillations, which are responsible for the observed QPOs. As the outburst evolves, the inward-propagating, relatively cool Keplerian disk enhances Compton cooling, causing the corona to contract. During the observations considered here, the source was in the SIMS. The observed emission is likely dominated by low-energy thermal photons directly from the disk and reflected photons produced by returning radiation in the strong gravitational field of the rapidly spinning black hole. A smaller contribution from Comptonized coronal photons and their reflected counterparts is also plausible.
Furthermore, polarimetric measurements provide an additional means of constraining the emission geometry and testing these radiation mechanisms. Following the disk- or slab-like geometry discussed by \citep{Podgorny26b}, we consider the combined polarization contributions from direct thermal disk emission, returning radiation reflected from the disk, and coronal emission and its reflection. Based on our preferred physical interpretation, Fig.~\ref{fig7} presents a schematic of the plausible accretion geometry and the associated radiation processes in connection to the spectro-temporal-polarimetric measurements \citep{Nandi2018}.

Future studies employing advanced physical polarimetric codes, such as \texttt{STOKES} \citep{Podgorny26b}, could further constrain the proposed emission geometry and distinguish between the contributions of direct disk emission, returning radiation, and coronal scattering. By incorporating complex radiative transfer effects, including multiple scattering, energy-dependent opacities, and general relativistic corrections, such studies could provide a more robust understanding of the inner accretion geometry and the physical mechanisms governing the observed polarization.

A brief summary of findings in this {\it paper} is as follows:

\begin{enumerate}[i)]

\item Comprehensive spectro-temporal-polarimetric analysis of GRS~1739--278 using observations from \textit{MAXI}/GSC, \textit{Swift}/XRT, \textit{NuSTAR} (FPMA/FPMB), \textit{IXPE}, and \textit{Insight}-HXMT (LE/ME) reveals that the $2025$--$2026$ outburst is a `failed' outburst.
The HID shows that recent outburst does not follow a canonical `q'-shaped track. Instead, the outburst is dominated by thermal soft X-ray photons, with a minimal contribution from high energy photons throughout the outburst. 

\item LFQPOs are observed in the low energy band ($2$--$4$~keV). These QPOs are consistent with `low-frequency' type-B QPOs observed during the soft intermediate spectral state.

\item The broadband spectral analysis indicates that the emission is dominated by the thermal soft X-ray component. Furthermore, the evolution of the component flux suggests significant broadening (or smearing) of the inner accretion disk near the outburst peak, which may produce thermal reflection features in the spectrum, consistent with a rapidly spinning BH.

\item A significant PD of $(2.43 \pm 0.47)$\% is detected in the total band $2$--$8$~keV  of {\it IXPE} at a PA of $(66.1 \pm 5.4)^\circ$ using \texttt{PCUBE} and a PD of $2.28 \pm 0.28$\% at a PA $(62.0 \pm 3.4)^\circ$ is estimated using \texttt{QUEEN-BEE}, showing a consistent measurement over different methods. 

\item Spectro-polarimetric analysis indicates the overall polarization due to direct disk emission, returning reflected photons, and a negligible non-thermal wedge-shaped corona component, with temporal analysis supporting a `weak' oscillating corona. Disk photons dominate at low energies, while the high polarization at $6$--$8$ keV likely originate from reflected photons.

\end{enumerate}

\section*{Acknowledgements}

The authors thank the anonymous referee for valuable suggestions that helped to improve the quality of the paper. 
This work made use of the archival data from NASA
observatories - \textit{Swift/XRT} and \textit{NuSTAR}, alongside \textit{IXPE}, a joint mission by NASA and Italian Space Agency (ASI). We also acknowledge the use of data from \textit{Insight-HXMT},  a satellite mission of the
China National Space Administration (CNSA) and on-demand data from \textit{MAXI/GSC}, a Japanese
Experiment Module on the International Space Station.
SS and HKC acknowledge the grants NSTC-114-2112-M-007-042 of the National Science and Technology Council (NSTC) of Taiwan. DD acknowledges the visiting research grant of National Tsing Hua University, Taiwan with the support from NSTC (NSTC-114-2811-M-007-084). 
AN thanks GH, SAG; DD, PDMSA, and Director, URSC for encouragement and continuous support to carry out this research.
We thank each instrument team for processing the data and providing the necessary software tools for the analysis.

\bibliography{references_grs1739m278}
\bibliographystyle{aasjournalv7.1}

\end{document}